\documentclass[10pt,a4paper]{article}

\usepackage{arxiv}
\usepackage{tikz}
\usepackage[utf8]{inputenc} 
\usepackage[T1]{fontenc}    
\usepackage{hyperref}       
\usepackage{url}            
\usepackage{booktabs}       
\usepackage{amsfonts}       
\usepackage{nicefrac}       
\usepackage{microtype}      
\usepackage{lipsum}

\usepackage{amssymb}
\usepackage{lineno,hyperref}
\usepackage{amsmath}
\usepackage{bm}
\usepackage{subcaption}
\usepackage{graphicx}
\usepackage{svg}
\usepackage{rotating}
\usepackage{lscape}
\usepackage{array,multirow}
\usepackage[title]{appendix}
\usepackage{float}
\usepackage{stmaryrd}
\usepackage{textcomp}
\usepackage{algorithm}
\usepackage{algpseudocode}
\usepackage{amsthm}
\usepackage{listings}
\usepackage{xcolor}
\usepackage{siunitx}

\usepackage{pifont}
\usepackage{lineno}

\usepackage{import}
\usepackage{physics}
\usepackage{comment}
\usetikzlibrary{positioning}

    \DeclareFontFamily{OT1}{pzc}{}
\DeclareFontShape{OT1}{pzc}{m}{it}{<-> s * [1.10] pzcmi7t}{}
\DeclareMathAlphabet{\mathpzc}{OT1}{pzc}{m}{it}
\usepackage{natbib}
\usepackage[algo2e,ruled,vlined]{algorithm2e}
\DontPrintSemicolon

\title{On physics-informed data-driven isotropic and anisotropic constitutive models through probabilistic machine learning and space-filling sampling}

\author{
  Jan Niklas Fuhg \\
  Sibley School of Mechanical and Aerospace Engineering \\
  Cornell University, 
   New York, USA \\
  \texttt{jf853@cornell.edu} \\

   \And
 Nikolaos Bouklas \\
  Sibley School of Mechanical and Aerospace Engineering\\
  Center for Applied Mathematics\\
  Cornell University,
   New York, USA \\
  \texttt{nb589@cornell.edu} \\
  }

\DeclareMathOperator*{\argmax}{arg\,max}

\begin{document}
\maketitle

\begin{abstract}
Data-driven constitutive modeling is an emerging field in computational solid mechanics with the prospect of significantly relieving the computational costs of hierarchical computational methods. Additionally, this data-driven paradigm could enable a seamless connection of experimental data probing material responses with numerical simulations at the structural level.
Traditionally, these surrogates have just been trained using datasets which map strain inputs to stress outputs directly. Data-driven constitutive models for elastic and inelastic materials have commonly been developed based on artificial neural networks (ANNs), which recently enabled the incorporation of physical laws in the construction of these models. However, ANNs do not offer convergence guarantees from an engineering point of view and are majorly reliant on user-specified parameters.
In contrast to ANNs, Gaussian process regression (GPR) is based on nonparametric modeling principles as well as on fundamental statistical knowledge and hence allows for strict convergence guarantees. GPR however has the major disadvantage that it scales poorly as datasets get large.
In this work we present a physics-informed data-driven constitutive modeling approach for isostropic and anisotropic materials at finite strain based on probabilistic machine learning that can be used in the big data context. 
This generalized approach is
based on rewriting the stress output as a linear combination of an irreducible integrity basis. The trained GPR surrogates are able to respect physical principles such as  material frame indifference, material symmetry, thermodynamic consistency, stress-free undeformed configuration, and the local balance of angular momentum.
Furthermore, this paper presents the first sampling approach that directly generates space-filling points in the invariant space corresponding to a bounded domain of the gradient deformation tensor. The sampling technique is based on simulated annealing and provides more efficient and reliable physics-informed constitutive models. 
Overall, the presented approach is tested on synthetic data from isotropic and anisotropic constitutive laws and
shows surprising accuracy even far beyond the limits of the training domain, indicating that the resulting surrogates can efficiently generalize as they incorporate knowledge about the underlying physics. 

\end{abstract}


\section{Introduction}
There has been an increased interest in machine learning (ML) tools in the computational sciences the last few years. 
This rise in popularity is due to multiple reasons: the ability of machine learning models to directly utilize experimental data in simulation environments, generalization capabilities of the machine learning tools, potential speed up in comparison to traditional numerical methods and their automatic differentiation framework.
For these reasons, machine learning tools have recently been used as a solution scheme for forward and inverse problems involving partial differential equations \citep{raissi2019physics, kadeethum2020physics, fuhg2021mixed, fuhg2021interval} or for the development of intrusive and non-intrusive reduced order modeling schemes for accelerated solutions of PDEs \citep{kadeethum2021non,hernandez2021deep, kadeethum2021framework}.
On the other hand the use of ML "black-box" models for constitutive modeling has been extensively studied for over 20 years. Starting from the influential works of \cite{wu1990representation,ghaboussi1990material,ghaboussi1991knowledge} for concrete in a small-strain biaxial state of stress, these tools have been employed for different material models with increasing complexity over the years \citep{lefik2003artificial,jung2006characterizing,huang2020machine,FUHG2021103522,lu2021stochastic,logarzo2021smart}.
Recently, in the hope of needing less data and in order to generate models with higher generalization capability, efforts have been made to train data-driven constitutive models that do not only train with raw stress-strain data but incorporate additional physics-based restrictions to the trained model
\citep{liu2019exploring,heider2020so,xu2021learning, linka2021constitutive}.
When dealing with hyperelastic materials, where no rate-dependence is considered, these models try to include some of the following physics-informed principles:  
\begin{itemize}
    \item Stress-free undeformed configuration: A rigid-body motion induces no strains and consequently no stresses.
    \item Material frame indifference: Tensor fields such as the stress and the strain should be objective under a change of observer.
    \item Material symmetry: Strain energy and stresses are consistent with existing symmetry groups in the material.
    \item Local balance of angular momentum: The Cauchy-stress tensor and the second Piola-Kirchhoff stress tensor should be symmetric.
    \item Thermodynamic consistency: Fulfillment of the Clausius-Planck inequality.
\end{itemize}

The idea behind physics-informed or physics-guided data-driven constitutive models is that the trained surrogate should abide to these conditions and not rely solely on raw data.
Additionally, a large majority of the proposed works in the literature for physics-guided constitutive models are based on artificial neural networks (ANNs) \citep{liu2019exploring,heider2020so,xu2021learning,masi2021thermodynamics}. For general information about ANNs we refer to \cite{goodfellow2016deep}.
However, ANNs have some characteristics that make them suboptimal with regards to training constitutive models from data:
\begin{enumerate}
    \item ANNs are parametric, which means that they are majorly dependent on user-specified parameter values such as hidden layers or the number of neurons per layer. Typically these are chosen based on user experience or in a grid-search fashion. However, this is widely viewed as one of the problematic aspects of ANNs. 
    \item Even though ANNs are known to be universal approximators \citep{hornik1989multilayer,lu2017expressive},
    this is only true if neural networks of either arbitrary width or arbitrary depths are available, which is not common for general engineering applications due to limited computational resources. Hence, there is a clear lack of convergence guarantees with regards to ANNs.
    \item ANNs do not have the ability to exactly represent points of the trained dataset due to (typically) being  defined as a mean-error minimizer. This is a significant concern for the preservation of the stress-free undeformed configuration.
\end{enumerate}
Due to these reasons, Gaussian process regression (GPR) also known as Kriging \citep{rasmussen2003gaussian} has recently gained more attention as a tool to fit constitutive data. The major factor for this is that in constrast to ANNs, GPR is based on nonparametric modeling principles as well as on fundamental statistical knowledge and hence allows for strict convergence guarantees and to obtain as output probabilistic information such as the mean and variance of the trained model.
Recently, the authors \citep{FUHG2021103522} have proposed a model-data-driven approach using GPR which enhances analytical constitutive models by local corrections based on data.
\cite{rocha2021fly} present a method relying on the adaptive construction of GPR models with application to elastoplastic multiscale mechanics.
\cite{wang2021metamodeling} use a method based on proper orthogonal decomposition (POD) and GPR known as POD-Kriging to build a surrogate for a time dependent constitutive model for a viscoelastic hydrogel.
One solution for probably the most significant problem associated with GPR in the big data regime has recently been proposed by the authors in \cite{fuhg2021local} where local approximate GPR (laGPR) has been presented for constitutive modeling applications and corresponding multiscale calculations.
\cite{frankel2019tensor} presented the only work so far where GPR and physics-guided data-driven constitutive modeling is attempted. They propose an approach for isotropic hyperelastic materials 
which is generally based on building a surrogate model that maps from the space of invariants of the right Cauchy-Green deformation tensor to the space of coefficients linked with the stress generators.
In this work we utilize the approach discussed in \cite{frankel2019tensor} as a starting point. We generalize it, extend it to anisotropic materials and by employing laGPR we allow the framework to be applicable in the big data context. By doing so we are able to capture five major physical constraints: the preservation of the stress-free undeformed configuration, material frame indifference, material symmetry conditions, thermodynamic consistency and local balance of angular momentum. 
Furthermore, we introduce  the first space-filling sampling approach which directly generates samples in principal and pseudo invariant space. Thereby, creating metamodels which prove to be more efficient.

The paper is structured as follows.
The general framework for modeling hyperelastic materials and the corresponding essential physical principles are introduced in Section \ref{sec::1}.
The physics-informed approach for hyperelastic data-driven constitutive laws for isotropic and anistropic materials are discussed and and explained in Section \ref{sec::3}.
A space-filling sampling approach in the invariant spaces of isotropic and anisotropic materials is introduced in Section \ref{sec::Sampling}. 
Local approximate Gaussian process regression as well as a consistent form to approximate the material tangent is presented in Section \ref{sec::2}.
The presented framework is studied on two numerical examples with different material symmetries in Section \ref{sec::4}.
The paper is concluded in Section \ref{sec::5}.

\section{Physics-based constraints for data-driven mappings}\label{sec::1}
\begin{figure}
    \centering
    \includegraphics[scale=0.5]{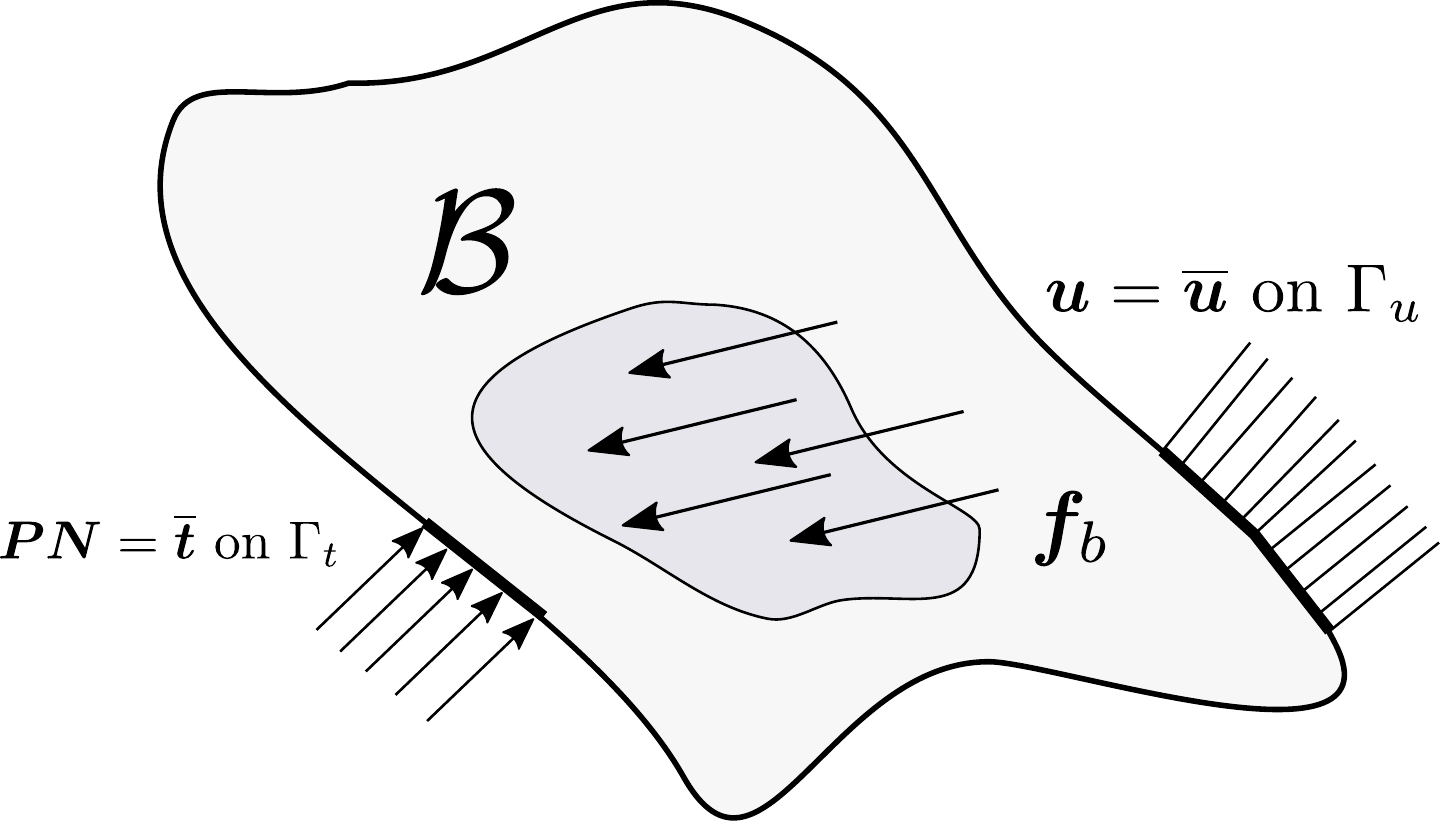}
    \caption{Solid domain with boundary conditions}
    \label{fig:pot}
\end{figure}

Consider an elastic body $\mathcal{B} \subset \mathbb{R}^{3}$. Let the boundary of the body $\Gamma$ be composed of two parts $\Gamma_{t}$ and $\Gamma_{u}$ such that $\Gamma = \Gamma_{t} \cup \Gamma_{u}$ (Fig. \ref{fig:pot}).
Here,  $\Gamma_{u}$ and  $\Gamma_{t}$ describe the boundary sections that displacement and traction boundary conditions are prescribed.
The time-dependent motion between the referential position $\bm{X}$ and the current position $\bm{x}$ can be defined by
\begin{equation}
\bm{x} = \bm{\varphi}(\bm{X},t) = \bm{X} + \bm{u}(\bm{X},t)
\end{equation} 
where $\bm{u}$ describes the time-dependent displacement field and $\bm{\varphi}(\bm{X},t)$ denotes the motion of the body. 
This allows to define the deformation gradient
\begin{equation}
\bm{F} = \text{Grad} \bm{\varphi}(\bm{X})
\end{equation}
and the right Cauchy-Green tensor
\begin{equation}
    \bm{C} = \bm{F}^{T} \bm{F}.
\end{equation}
From the balance of linear momentum, the local equilibrium equation
\begin{equation}
    \text{div} \bm{\sigma} + \bm{b} = \bm{0}
\end{equation}
is fulfilled where $ \bm{\sigma} $ is the Cauchy stress tensor and $ \bm{b} $ are body forces. The balance of angular momentum leads to the symmetry of the Cauchy stress tensor
\begin{equation}
    \bm{\sigma} = \bm{\sigma}^{T}.
\end{equation}
Furthermore, knowing that $\bm{S} = \sqrt{\text{det}(\bm{C})} \bm{F}^{-1} \bm{\sigma} \bm{F}^{-T}$ \citep{holzapfel2000nonlinear}, where $\bm{S}$ is the second Piola-Kirchhoff stress tensor, we can see that $\bm{S} = \bm{S}^{T}$ must hold as well. Hence, any data-driven constitutive law with $\bm{S}$ as an output must ensure its symmetry. \\

In the hyperelastic framework, the existence of the strain energy function $\Psi$ is postulated, which is assumed to be defined per unit reference volume \citep{holzapfel2000nonlinear}.
The formulation of an explicit strain energy function is dependent on the symmetry group that complies with the symmetries that correspond to a specific material. 
A symmetry group of a material is a set of transformations that allow for material symmetry to be preserved.
The three dimensional orthogonal group is $\bm{O}\text{rth}$ is defined as the group of $3×\times3$ orthogonal matrices
\begin{equation}
    \bm{O}\text{rth} = \lbrace \bm{R} \in \mathbb{R}^{3} \otimes \mathbb{R}^{3} | \bm{R}^{T} = \bm{R}^{-1} \rbrace.
\end{equation}
Let there be some some structural tensors $\bm{A}_{i}$, $i=1,\ldots p$ that determine the symmetry group $\mathcal{G}$  of an anisotropic material with
\begin{equation}\label{eq:SymGroup}
    \mathcal{G} = \lbrace \bm{R} \in \bm{O}\text{rth} | \bm{R}^{T} \bm{A}_{i} \bm{R}, \forall i=1, 
    \ldots,p \rbrace.
\end{equation}
According to \cite{ehret2007polyconvex} a convenient way to describe the structural tensors is by defining them as
\begin{equation}
    \bm{A}_{i} = \bm{a}_{i} \otimes \bm{a}_{i}, \qquad i=1, \ldots p
\end{equation}
where the $\bm{a}_{i}\in \mathbb{R}^{3}$ are  unit vectors.
For a general material consider the strain energy density to be at least a function of the Cauchy-Green tensor 
\begin{equation}
    \Psi = \Psi(\bm{C}, \bullet).
\end{equation}
Under this assumption a necessary condition for material symmetry in terms of eq. (\ref{eq:SymGroup}) is given by
\begin{equation}\label{eq::FirstSym}
    \Psi(\bm{R}^{T} \bm{C} \bm{R}) = \Psi(\bm{C}), \qquad \forall \bm{R} \in \mathcal{G}.
\end{equation}
Following
\cite{zhang1990structural} and \cite{itskov2004class}, this condition can only   be ensured if the strain energy density function has the structural tensors as additional arguments, i.e. $\Psi(\bm{C},\bm{A}_{i})$, $i=1, \ldots,p$.
With regards to the symmetry group $\mathcal{G}$ it is therefore necessary to require
\begin{equation}\label{eq::finalSy}
    \Psi(\bm{R}^{T} \bm{C} \bm{R}, \bm{R}^{T} \bm{A}_{i} \bm{R}) = \Psi(\bm{C},\bm{A}_{i}), \qquad i=1, \ldots,p, \qquad \forall \bm{R} \in \mathcal{G}
\end{equation}
for material symmetry. Data-driven constitutive laws should therefore aim to satisfy eq. (\ref{eq::finalSy}).
Another important requirement for data-driven material models is material frame indifference, i.e. independence of the observer, which reads
\begin{equation}\label{eq::finalMaterFrame}
    \Psi(\bm{R}^{T} \bm{C} \bm{R}, \bm{R}^{T} \bm{A}_{i}  \bm{R}) =\bm{R}^{T} \Psi(\bm{C},\bm{A}_{i}) \bm{R}, \qquad i=1, \ldots,p, \qquad \forall \bm{R} \in \bm{O}\text{rth}.
\end{equation}
Even though it is not a theoretical requirement, hyperleastic constitutive laws are constructed so that they satisfy that the reference configuration corresponds to a stress-free state with zero strain energy density. This can be achieved by requiring that
\begin{equation}
    \Psi(\bm{I}, \bullet) = 0
\end{equation}
in the reference configuration where $\bm{F} = \bm{C} =\bm{I}$. This condition and the physically sound assumption that the strain energy function increases under deformation, i.e. $ \Psi(\bm{C}, \bullet) \geq 0$, ensures that the stress in the reference configuration is zero, i.e.
\begin{equation}
    \bm{S}(\bm{I}, \bullet) = \bm{0}
\end{equation}
which is another physical constraint for data-driven material models. 
 
If a scalar-valued tensor function, such as the strain energy density function, is invariant under a rotation it may written in terms of the invariants ($I_{1}, \ldots, I_{J}$) of its arguments \citep{holzapfel2000nonlinear}.
For purely mechanical processes of perfectly elastic materials, the second law of thermodynamics requires that the Clausius-Planck inequality turns into an equality, satisfying
\begin{equation}
    \bm{S}: \dot{\bm{E}} - \dot{\Psi} =0
\end{equation}
which with arbitrary strain measures requires that
\begin{equation}\label{eq::StressTens}
\begin{aligned}
\bm{S} &=2\frac{\partial \Psi(\bm{C},\bm{A}_{1}, \ldots \bm{A}_{p})}{\partial \bm{C}}\\ &=  2\frac{\partial \Psi(I_{1}, \ldots, I_{J})}{\partial \bm{C}} = 2 \sum_{i=1}^{J} \frac{\partial \Psi}{\partial I_{i}} \frac{\partial I_{i}}{\partial \bm{C}}.
\end{aligned}
\end{equation}
Hence, any data-driven constitutive model that fulfills this relationship is thermodynamically consistent, at least for the case of perfectly elastic materials.
From the last term of eq. (\ref{eq::StressTens}) it can be seen that the stress can be written as a linear combination of some tensors $\frac{\partial I_{i}}{\partial \bm{C}}$. 
These terms can be condensed to build an integrity basis $\mathbb{G}$ for the stress, constructed from some symmetric components $\bm{H}_{i} \in \mathbb{R}^{3 \times 3}$, $i=1, \ldots, D$, as
\begin{equation}
    \mathbb{G} = \lbrace \bm{H}_{1}, \ldots \bm{H}_{D} \rbrace.
\end{equation}

This in turn allows us to write the functional mapping of the stress $\bm{S} = \Phi(\bm{C},\bm{A}_{1}, \ldots \bm{A}_{p})$ as a linear combination of the components of the integrity basis as
\begin{equation}\label{eq::StressTensWrittem}
    \bm{S} = c_{1}(I_{1}, \ldots, I_{J}) \bm{H}_{1}(\bm{C},\bm{A}_{1}, \ldots \bm{A}_{p}) + \ldots +c_{D}(I_{1}, \ldots, I_{J}) \bm{H}_{D}(\bm{C},\bm{A}_{1}, \ldots \bm{A}_{p}).
\end{equation}
If none of the components of $\mathbb{G}$ is expressible as a linear combination of the others this representation is called an irreducible representation. 
Since we already assumed that the strain energy function is material frame indifferent it can be shown that any representation of the form of eq. (\ref{eq::StressTensWrittem}) fulfills the condition of material frame indifference as well \citep{zheng1994theory}. Additionally, by construction any stress generated by eq. (\ref{eq::StressTensWrittem}) respects the material symmetry condition of eq. (\ref{eq::finalSy}) \citep{zheng1994theory,xiao1995general}.
One problem with this approach is that an irreducible integrity basis has to be known.
However, in the last 50 years a lot of effort has been put into finding suitable irreducible function basis for different kinds of material anisotropies.
In particular, \cite{zheng1994theory} presented a unified invariant approach for the representation of tensor functions for different cases of anisotropy (isotropy, hemitropy, transversal isotropy, orthotropy) in two and three dimensions. 
The findings of this work have been extensively studied and applied
\citep{schroder2003invariant,itskov2004class, balzani2006polyconvex}.
Therefore, the integrity bases for standard anisotropic cases are known. \\

Now consider a material model dataset given by strain inputs and stress outputs of the form
\begin{equation}
    \mathcal{D} = \lbrace \bm{C}^{i}, \bm{S}^{i} \rbrace_{i=1}^{N}
\end{equation}
and assume that the material anisotropy and corresponding structural tensors $(\bm{A}_{1}, \ldots \bm{A}_{p})$ are known. This allows us to obtain the relevant invariants for every input ($I_{1}^{i}, \ldots, I_{J}^{i}$). Furthermore assume that by knowing the $\bm{C}_{i}$ and  $\bm{S}_{i}$ the respective coefficients ($c_{1}^{i}, \ldots, c_{D}^{i}$) of eq. (\ref{eq::StressTensWrittem}) can be obtained.
Hence, we can generate the alternative dataset
\begin{equation}
    \mathcal{D}_{alt} = \lbrace [I_{1}^{i}, \ldots, I_{J}^{i}], [c_{1}^{i}, \ldots, c_{D}^{i}] \rbrace_{i=1}^{N}
\end{equation}
and build a surrogate model 
\begin{equation}\label{eq::GeneralMapping}
    \hat{\mathcal{M}}: \mathcal{I} \in \mathbb{R}^{J} \rightarrow \mathpzc{c} \in \mathbb{R}^{D}
\end{equation}
for this dataset. This metamodel allows us to obtain an approximation of the stress output with
\begin{equation}\label{eq::estimateStress}
   \hat{\bm{S}}^{i} = \hat{c}^{i}_{1}(I_{1}^{i}, \ldots, I_{J}^{i}) \bm{H}^{i}_{1} + \ldots + \hat{c}^{i}_{D}(I_{1}^{i}, \ldots, I_{J}^{i}) \bm{H}^{i}_{D}
\end{equation}
where $\hat{c}^{i}_{j}$ refers to the $j$-th output of the trained metamodel for the $i$-th input.
Here, the symbol $\hat{\bullet}$ indicates an approximated value.

Following the mapping approach of eq. (\ref{eq::GeneralMapping}) we can easily incorporate the following physics-based constraints into a data-driven constitutive model:
\begin{itemize}
    \item Local balance of angular momentum: Because the integrity basis is necessarily symmetric,
    \item Material frame indifference : By estimating stress with eq. (\ref{eq::estimateStress}),
    \item Material symmetry conditions: By design the stress output of eq. (\ref{eq::estimateStress}) respects the symmetry of isotropic and anisotropic materials,
    \item Stress-free undeformed configuration: When one set of input invariants are chosen that correspond to the undeformed configuration and a surrogate modeling technique with exact inference (such as GPR) is used,
    \item Thermodynamic consistency: By approximating the stress with eq. (\ref{eq::estimateStress}) thermodynamic consistency holds at the training points when a metamodeling method with exact inference properties is used.
\end{itemize}
In the following we explicitly highlight the details of this mapping approach for isotropic and transversally isotropic materials and compare it to the predominant approach of training a metamodel through strain and stress data.

\section{Physics-informed mapping for hyperelastic materials}\label{sec::3}
In this section we discuss and highlight different variations of data-preprocessing for hyperelastic laws.
\subsection{Classical mapping approach}
The classical mapping approach, see e.g. \cite{ghaboussi1998autoprogressive,hashash2004numerical,lefik2009artificial}, is based on the symmetry condition of the right Cauchy-Green tensor and the second Piola-Kirchhoff stress. This allows to postulate a mapping between upper triangular components of the two tensors, i.e.
\begin{equation}\label{eq:naiveMap}
\begin{bmatrix}
C_{11} \\ C_{12} \\ C_{13} \\ C_{22} \\ C_{23} \\ C_{33}
\end{bmatrix}
\rightarrow
\begin{bmatrix}
S_{11} \\ S_{12} \\ S_{13} \\ S_{22} \\ S_{23} \\ S_{33}
\end{bmatrix}.
\end{equation}
This principle can equivalently be applied using other symmetric strain and stress measures, e.g. engineering strain and Cauchy stress. If the data is only available in a non-symmetric tensor form, for example as the deformation gradient and the first Piola-Kirchhoff stress, then they can simply be converted to equivalent symmetric tensors. 
The mapping of equation (\ref{eq:naiveMap}) is independently utilizable regardless of any type of anisotropy implicitly present in the data and allows to easily capture two physics informed-principles: local balance of angular momentum and preservation of the stress-free undeformed configuration.
\subsection{Physics-informed mapping approach for isotropic materials}
For the isotropic case we follow we the approach proposed by \cite{frankel2019tensor}.
The isotropic case is fully defined by the 3 invariants
\begin{equation}
\begin{aligned}
I_{1} &= \text{tr}(\bm{C}) \\
I_{2} &= 0.5 (\text{tr}(\bm{C})^{2} - \text{tr}(\bm{C}^{2})) \\
I_{3} &= \text{det}(\bm{C}) .
\end{aligned}
\end{equation}
The second Piola-Kirchhoff stress response of an isotropic material can always be decomposed into the three stress generators \citep{holzapfel2000nonlinear}
\begin{equation}\label{eq::IsoStressGenerator}
    \mathbb{G}  =\lbrace \bm{I} , \bm{C}, \bm{C}^{-1} \rbrace 
\end{equation}
with
\begin{equation}\label{eq::IsoMapping}
\bm{S} = c_{1}(I_{1},I_{2},I_{3}) \bm{I} + c_{2}(I_{1},I_{2},I_{3})  \bm{C} + c_{3}(I_{1},I_{2},I_{3})  \bm{C}^{-1}.
\end{equation}
Therefore instead of learning a mapping between the symmetric components of $\bm{C}$ and $\bm{S}$ as described in equation (\ref{eq:naiveMap}), we can learn a functional mapping of the form
\begin{equation}
\begin{bmatrix}
I_{1} \\ I_{2} \\ I_{3}
\end{bmatrix}
\rightarrow
\begin{bmatrix}
c_{1} \\ c_{2} \\ c_{3}
\end{bmatrix}
.
\end{equation}
Hence, instead of mapping $\mathbb{R}^{6} \rightarrow \mathbb{R}^{6}$, we map $\mathbb{R}^{3} \rightarrow \mathbb{R}^{3}$.
Since all the stress generators of equation (\ref{eq::IsoStressGenerator}) are symmetric, any output of equation (\ref{eq::IsoMapping}) is symmetric as well, and hence, the local balance of angular momentum is ensured.
 Furthermore, the presented mapping yields an output which necessitates the fulfillment of material symmetry and material frame indifference.
 To prove the latter, consider the rotation tensor $\bm{R}$, then we see that the following needs to hold
 \begin{equation}
\bm{S}' = \bm{R} \bm{S} \bm{R}^{T} = \bm{R} \Phi(\bm{C}) \bm{R}^{T} =  \Phi(\bm{R} \bm{C} \bm{R}^{T}) .
\end{equation}
This can be proven by using the tensor generators
\begin{equation}
\begin{aligned}
\bm{S}' &= \bm{R} \bm{S} \bm{R}^{T} \\
&= \bm{R} (c_{1} \bm{I} + c_{2} \bm{C} + c_{3}\bm{C}^{-1}) \bm{R}^{T} \\
&= c_{1} \bm{R}\bm{R}^{T}  + c_{2} \bm{R}\bm{C}\bm{R}^{T} + c_{3}\bm{R} \bm{C}^{-1}\bm{R}^{T} \\
&=  \Phi(\bm{R} \bm{C} \bm{R}^{T}) .
\end{aligned}
\end{equation}
The scalar values $c_{1}, c_{2}, c_{3}$ can be obtained by observing that
\begin{equation}
    \bm{Q}\bm{S}\bm{Q}^{T} = \begin{bmatrix}
    \lambda_{1}^{s} & 0 & 0 \\
    0 & \lambda_{2}^{s} & 0 \\
    0 & 0 & \lambda_{3}^{s}
    \end{bmatrix},
    \qquad
       \bm{Q}\bm{C}\bm{Q}^{T} = \begin{bmatrix}
    \lambda_{1}^{C} & 0 & 0 \\
    0 & \lambda_{2}^{C} & 0 \\
    0 & 0 & \lambda_{3}^{C}
    \end{bmatrix}
\end{equation}
which following \cite{frankel2019tensor} allows to define an equation system for the unknown scalar values of the form
\begin{equation}
 \begin{bmatrix}
    c_{1} \\
    c_{2} \\
    c_{3}
    \end{bmatrix}
    = \begin{bmatrix}
    1 & \lambda_{1}^{C} & \frac{1}{\lambda_{1}^{C}} \\
    1 & \lambda_{2}^{C} & \frac{1}{\lambda_{2}^{C}} \\
    1 & \lambda_{3}^{C} & \frac{1}{\lambda_{3}^{C}} \\
    \end{bmatrix}^{-1}
    \begin{bmatrix}
    \lambda_{1}^{s} \\
    \lambda_{2}^{s} \\
    \lambda_{3}^{s}
    \end{bmatrix}.
\end{equation}
The matrix might be severely ill-conditioned depending on the number of unique principal strains. \cite{frankel2019tensor} describe an algorithmic way to avoid this problem. \\
The consistent material tangent of the general stress formulation of eq. (\ref{eq::IsoMapping}) is given by

\begin{equation}\label{eq::IsoConsistent}
\begin{aligned}
  \mathbb{C} &= 2 \frac{\partial \bm{S}}{ \partial \bm{C}} \\
  &= 2 \left(\frac{\partial c_{1}}{ \partial \bm{C}} \otimes \bm{I} +c_{1} \frac{\partial  \bm{I}}{ \partial \bm{C}} + \frac{\partial c_{2}}{ \partial \bm{C}} \otimes \bm{C} +c_{2} \frac{\partial  \bm{C}}{ \partial \bm{C}} + \frac{\partial c_{3}}{ \partial \bm{C}} \otimes \bm{C}^{-1} +c_{3} \frac{\partial  \bm{C}^{-1}}{ \partial \bm{C}}\right)
\end{aligned}
\end{equation}
where
\begin{equation}
    \begin{aligned}
        \frac{\partial c_{i}}{ \partial \bm{C}} &= 
       &= \frac{\partial c_{i}}{ \partial I_{1}} \bm{I} + \frac{\partial c_{i}}{ \partial I_{2}} ( I_{1}\bm{I} - \bm{C}) + \frac{\partial c_{i}}{ \partial I_{3}} I_{3} \bm{C}^{-1}.
    \end{aligned}
\end{equation}
All remaining unknown derivatives of eq. (\ref{eq::IsoConsistent}) can be found in eq. (\ref{eq::ListOfGradientExpressions}).
Additionally, from eq. (\ref{eq::IsoConsistent}) it can be seen that if a surrogate model is trained taking as input the principal invariants of $\bm{C}$ and as output the scalar coefficients of  eq. (\ref{eq::IsoMapping}), taking the output derivative with regards to the input $(\frac{\partial c_{i}}{ \partial I_{j}})$ allows us to obtain an approximation of the consistent material tangent.  Both ANNs and GPR have the ability to obtain these derivatives. Other surrogate modeling techniques might need to rely on numerical differentiation, e.g. in the form of finite difference schemes \citep{miehe1996numerical}.

\subsection{Physics-informed mapping approach for transversly isotropic materials}
Transverse isotropy is characterized by a single unit direction $\bm{a}_{0}$ that characterizes the material symmetries. Hence, only one structural tensor $\bm{A} =\bm{a}_{0} \otimes \bm{a}_{0} $ is needed to fulfill the material symmetry of the strain energy function.
The principal invariants of the right Cauchy-Green tensor read
\begin{equation}\label{eq::PrincipalInvariants}
\begin{aligned}
I_{1} &= \text{tr}(\bm{C}) \\
I_{2} &= 0.5 (\text{tr}(\bm{C})^{2} - \text{tr}(\bm{C}^{2})) \\
I_{3} &= \text{det}(\bm{C}) .
\end{aligned}
\end{equation}
We furthermore need to consider the two independent components of the pseudo invariants
\begin{equation}\label{eq::MixedInvariants}
\begin{aligned}
I_{4} &= \text{tr}(\bm{A} \bm{C}) \\
I_{5} &= \text{tr}(\bm{A} \bm{C}^{2}).
\end{aligned}
\end{equation}
Following \cite{zheng1994theory} the stress output can be decomposed into the six generators collected in the set $\mathbb{G}$
\begin{equation}
    \mathbb{G} = \lbrace \bm{I}, \bm{C}, \bm{A},\bm{C}^{2},(\bm{A}\bm{C} + \bm{C}\bm{A}), (\bm{A}\bm{C}^{2} + \bm{C}^{2}\bm{A})  \rbrace
\end{equation}
with
\begin{equation}\label{eq:transIso}
\bm{S} = c_{1} \bm{I} + c_{2} \bm{C} + c_{3} \bm{A} + c_{4} \bm{C}^{2} + c_{5} (\bm{A}\bm{C} + \bm{C}\bm{A}) + c_{6} (\bm{A}\bm{C}^{2} + \bm{C}^{2}\bm{A}).
\end{equation}
Therefore, in the transversely isotropic case we can learn a function mapping from
\begin{equation}\label{eq:transIMapping}
\begin{bmatrix}
I_{1} \\ I_{2} \\ I_{3}\\ I_{4}\\ I_{5}
\end{bmatrix}
\rightarrow
\begin{bmatrix}
c_{1} \\ c_{2} \\ c_{3}\\ c_{4}\\ c_{5}\\ c_{6}
\end{bmatrix}.
\end{equation}
 So instead of mapping $\mathbb{R}^{6} \rightarrow \mathbb{R}^{6}$ as in the classical mapping case, we map $\mathbb{R}^{5} \rightarrow \mathbb{R}^{6}$ and are able to ensure local balance of angular momentum (all stress generators are symmetric) and maintain material symmetry conditions, thermodynamic consistency and material frame indifference.
The latter can easily be proven by considering
\begin{equation}
\bm{S}' = \bm{R} \bm{S} \bm{R}^{T} = \bm{R} \Phi(\bm{C}, \bm{A}) \bm{R}^{T} =  \Phi(\bm{R} \bm{C} \bm{R}^{T}, \bm{R} \bm{A} \bm{R}^{T}) 
\end{equation}
and
\begin{equation}
\begin{aligned}
\bm{S}' &= \bm{R} \bm{S} \bm{R}^{T} \\
& =\bm{R} \left( c_{1} \bm{I} + c_{2} \bm{C} + c_{3} \bm{A} + c_{4} \bm{C^{2}} + c_{5} (\bm{A}\bm{C} + \bm{C}\bm{A}) + c_{6} (\bm{A}\bm{C}^{2} + \bm{C}^{2}\bm{A}) \right) \bm{R}^{T}\\
&= c_{1} \bm{R}\bm{R}^{T} + c_{2} \bm{R}\bm{C}\bm{R}^{T}  + c_{3} \bm{R}\bm{A} \bm{R}^{T}+ c_{4} \bm{R}\bm{C}^{T}\bm{R}^{T}\bm{R}\bm{C}\bm{R}^{T} \\
& + c_{5} (\bm{R} \bm{A}\bm{R}^{T}\bm{R} \bm{C}\bm{R}^{T} +\bm{R} \bm{C}\bm{R}^{T}\bm{R}\bm{A}\bm{R}^{T}) \\
&+ c_{6} (\bm{R}\bm{A}\bm{R}^{T}\bm{R}\bm{C}^{T} \bm{R}^{T}\bm{R}\bm{C}\bm{R}^{T} +\bm{R} \bm{C}^{T}\bm{R}^{T}\bm{R} \bm{C}\bm{R}^{T}\bm{R}\bm{A}\bm{R}^{T}) \\
&=  \Phi(\bm{R} \bm{C} \bm{R}^{T}, \bm{R} \bm{A} \bm{R}^{T}) .
\end{aligned}
\end{equation}
Obtaining the scalars for the isotropic case was straightforward using prinicpal spaces. However, for the transversely anisotropic case this would result in an underdetermined equation system.
Hence, given a stress output and a corresponding right Cauchy-Green tensor value the scalar values $c_{1}, \ldots, c_{6}$ can instead be approximated using a least squares approach. By rewriting equation (\ref{eq:transIso}) we get the overdetermined equation system
\begin{equation}
    \begin{aligned}
    \underbrace{
        \begin{bmatrix}
        \text{vec}(\bm{S}) \end{bmatrix}}_{\bm{b}} = \underbrace{\begin{bmatrix}
        \text{vec}(\bm{I}) & \text{vec}(\bm{C}) &\text{vec}(\bm{A}) &\text{vec}(\bm{C}^{2}) & \text{vec}(\bm{A}\bm{C}+\bm{C}\bm{A}) & \text{vec}(\bm{A}\bm{C}^{2}+\bm{C}^{2}\bm{A})
        \end{bmatrix}}_{\bm{A}}
        \underbrace{
\begin{bmatrix}
        c_{1} \\
        c_{2} \\
        c_{3} \\
        c_{4} \\
        c_{5} \\
        c_{6}
        \end{bmatrix}}_{\bm{x}}
    \end{aligned}.
\end{equation}
By defining the following optimization problem we can constrain the solution space to the linear least squares solution
\begin{equation}\label{eq::TransEquationScalar}
    \min_{x} \qquad \norm{\bm{A}\bm{x}-\bm{b}}_{2}^{2}
\end{equation}
which can for example be solved with QR-decomposition.
As a sidenote, instead of setting up a fully-determined equation system with six symmetric components of equation (\ref{eq:transIso}), we found that adding all $9$ possible equations makes the solution algorithm more robust. \\

Similarly to the isotopic case, the consistent material tangent for the transversally isotropic case can also be derived analytically and is only dependent on an approximation of the derivatives $(\frac{\partial c_{i}}{ \partial I_{j}})$ which can be obtained straightforwardly with some surrogate modeling techniques. The tangent is given as
\begin{equation}\label{eq::TransConsistentTangent}
\begin{aligned}
  \mathbb{C} &= 2 \frac{\partial \bm{S}}{ \partial \bm{C}} \\
  &= 2 \left(\frac{\partial c_{1}}{ \partial \bm{C}} \otimes \bm{I} +c_{1} \frac{\partial  \bm{I}}{ \partial \bm{C}} + \frac{\partial c_{2}}{ \partial \bm{C}} \otimes \bm{C} +c_{2} \frac{\partial  \bm{C}}{ \partial \bm{C}} + \frac{\partial c_{3}}{ \partial \bm{C}} \otimes \bm{A} +c_{3} \frac{\partial  \bm{A}}{ \partial \bm{C}}
  + \frac{\partial c_{4}}{ \partial \bm{C}} \otimes \bm{C}^{2} +c_{4} \frac{\partial  \bm{C}^{2}}{ \partial \bm{C}}
  \right. \\
  &+\left.  
  \frac{\partial c_{5}}{ \partial \bm{C}} \otimes (\bm{A}\bm{C} + \bm{C}\bm{A}) +c_{5} \frac{\partial  (\bm{A}\bm{C} + \bm{C}\bm{A})}{ \partial \bm{C}} 
  +
  \frac{\partial c_{6}}{ \partial \bm{C}} \otimes (\bm{A}\bm{C}^{2} + \bm{C}^{2}\bm{A}) +c_{6} \frac{\partial  (\bm{A}\bm{C}^{2} + \bm{C}^{2}\bm{A})}{ \partial \bm{C}} 
  \right)
      \end{aligned}
\end{equation}
where 
\begin{equation}
    \begin{aligned}
        \frac{\partial c_{i}}{ \partial \bm{C}} &= 
       &= \frac{\partial c_{i}}{ \partial I_{1}} \bm{I} + \frac{\partial c_{i}}{ \partial I_{2}} ( I_{1}\bm{I} - \bm{C}) + \frac{\partial c_{i}}{ \partial I_{3}} I_{3} \bm{C}^{-1} + \frac{\partial c_{i}}{ \partial I_{4}} \bm{A} + \frac{\partial c_{i}}{ \partial I_{5}} (\bm{a}_{0} \otimes \bm{C} \bm{a}_{0} +  \bm{a}_{0} \bm{C} \otimes \bm{a}_{0}).
    \end{aligned}
\end{equation}
All unknown tensor derivatives of eq. (\ref{eq::TransConsistentTangent}) are listed in eq. (\ref{eq::ListOfGradientExpressions}).\\
As a sidenote, orthotropic materials are not explicitly discussed in this paper. However they have seven invariants and seven generators \citep{zheng1994theory}. Hence, the scalar values of orthotropic materials can be obtained in a similar manner to equation (\ref{eq::TransEquationScalar}). After this the setup of the data-driven constitutive model approach for this anisotropic material class is straightforward. The next section proposes an approach of sampling new points in the invariant space.\\

\section{Space-filling sampling approach in invariant space}\label{sec::Sampling}
A problem associated with training a mapping where inputs are principal and pseudo invariants of the right Cauchy-Green deformation tensor, is that the sample placement in this space might not be spaced evenly even when sampling the deformation tensors in a space-filling way. Furthermore, generating samples in a space-filling fashion in a specific domain of the invariant space has not been sufficiently explored since it is not clear how to define a relevant region. 
This is important in order to evaluate how much trust we put into the output of a metamodel, i.e. if we have sampled points inside a restricted domain of the deformation gradient space and we train our surrogate with these points we expect the model to predict the correct output somewhat accurately if the new input point is inside the bounded training domain. However, outside this training domain we should not blindly trust the predicted surrogate output.
Hence, when using points in the invariant space as the model input we need to be able to understand which bounded domain of the deformation gradient space we are representing with our input data in order to judge how trustful a trained model is with regards to a certain deformation gradient input as is needed when employing the trained model in a FEM framework.\\

To highlight all these points, consider that a trained model should be able to accurately predict the constitutive law when the bounds of the deformation gradient components are given by
\begin{equation}\label{eq::DefoSpace}
        \overline{F}_{ij} \in [F^{L}_{ij}, F^{U}_{ij}] \text{ where } \begin{cases}
        1-\delta \leq 1 \leq 1+\delta, & \text{when } i=j \\
        -\delta \leq 0 \leq \delta, & \text{when } i\neq j
        \end{cases}.
\end{equation}
with $\delta>0$. This defines a nine-dimensional bounded space that samples need to be generated in.
In order to generate distributed points in a bounded space in computational engineering applications, latin hyercube sampling (LHS) \citep{stein1987large} or some form of optimal latin hypercube sampling (e.g. latin hypercube samples obtained with a translational propagation algorithm (TPLHD) \citep{viana2010algorithm})  are typically applied.
In order to highlight the problems associated with generating evenly-spaced samples in the deformation gradient space, i.e. in the bounds of eq. (\ref{eq::DefoSpace}), and transferring these samples into principal or pseudo invariant space (using eq. (\ref{eq::PrincipalInvariants}) or eq. (\ref{eq::MixedInvariants}) equivalently) assume in the following that $\delta=0.175$. Further on, we will refer to this training domain as the $17.5\%$ input domain. \\

\begin{figure}[h!]
    \centering
    \includegraphics[scale=0.5]{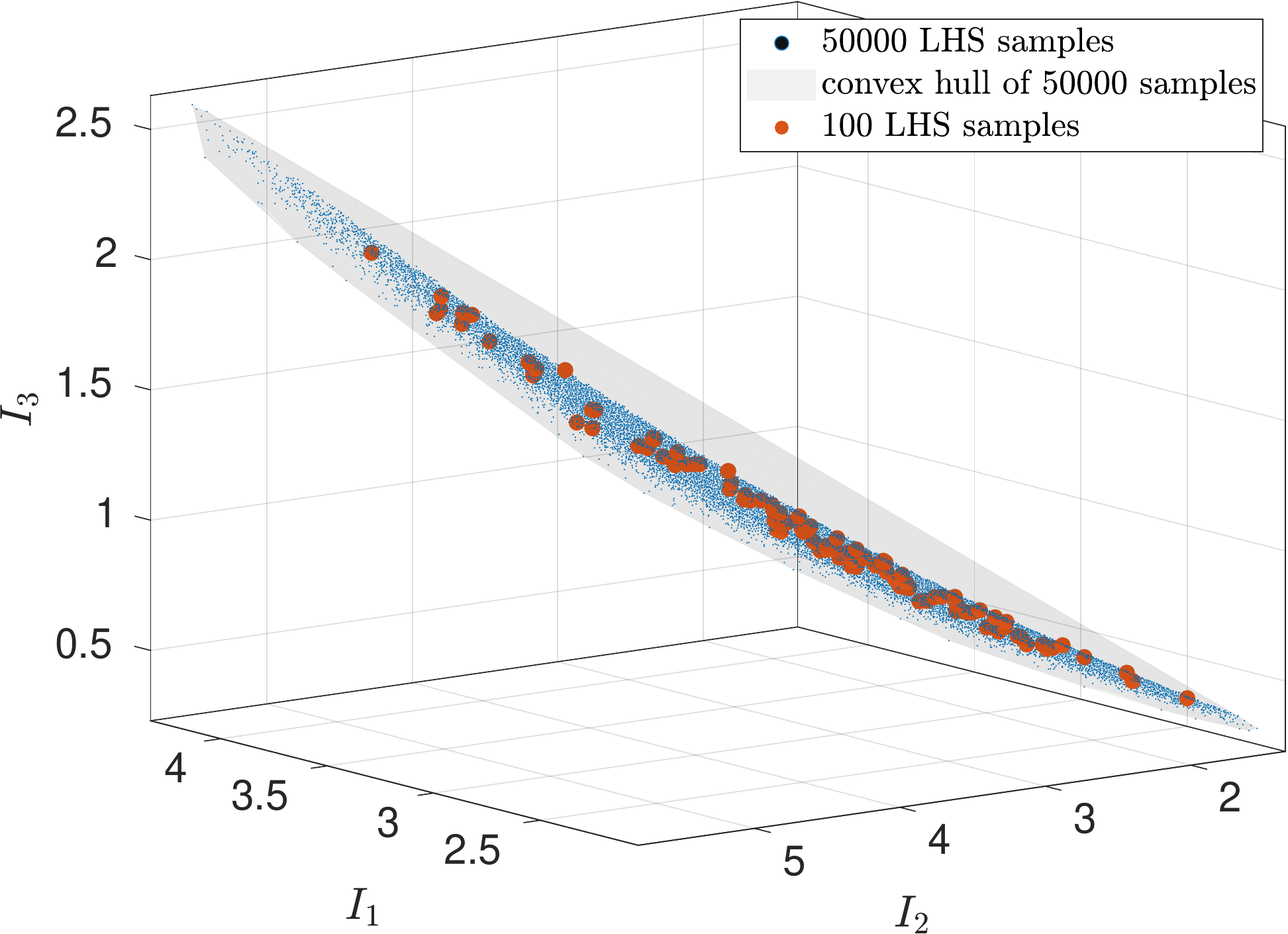}
    \caption{Example of unevenly spread samples in principal invariant space when mapping from the space of the deformation gradient space which was evenly-sampled with LHS in the $17.5\%$ input domain. }
    \label{fig:exampleWhyNeeded}
\end{figure}

\begin{figure}[h!]
    \centering
    \includegraphics[scale=0.5]{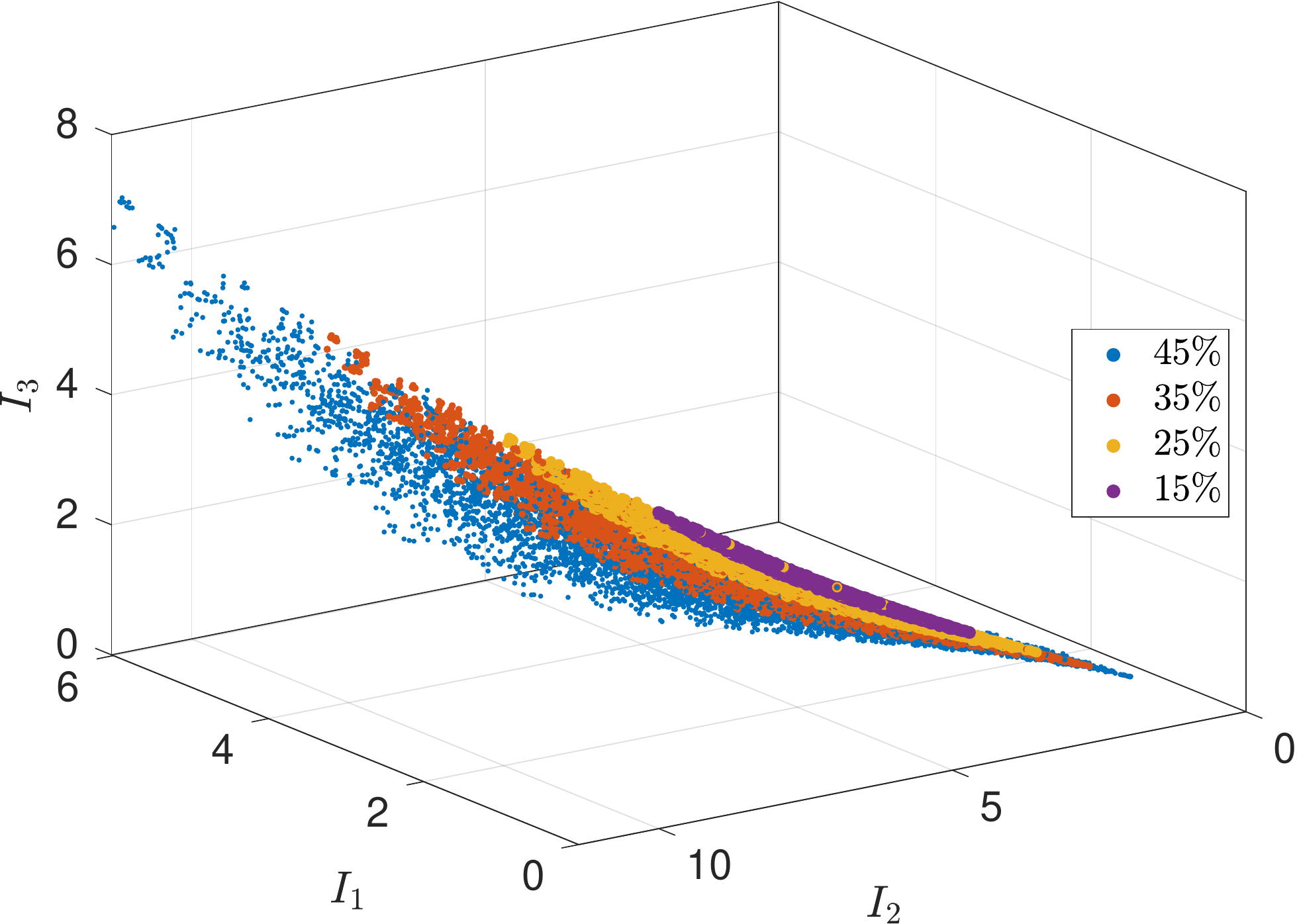}
    \caption{Spread samples in principal invariant space when mapped from $20,000$ samples generated in deformation gradient space with different $\delta$ bounds according to eq. (\ref{eq::DefoSpace}). }
    \label{fig:increasingSpredInPrincicap}
\end{figure}

Next, we want to visually inspect the bounded domain that the $17.5\%$ input region in deformation gradient space corresponds to in the principal invariant space. For this task we generate $50,000$ samples in the bounds of eq. (\ref{eq::DefoSpace}) with LHS and plot them in a scatter plot, see blue dots in Figure \ref{fig:exampleWhyNeeded}.
It can be seen that the effective area in invariant space  stretches over a narrow 3D band. Next, we  generate $100$ samples in the deformation gradient space with LHS and transfer them into invariant space, see red dots in Figure \ref{fig:exampleWhyNeeded}. We can observe that the resulting points are  unevenly distributed which will likely result in poor surrogate model performances in areas with low sample point density.
There is one additional problem associated with sampling in deformation gradient space as the mapping between the two spaces is not injective, i.e. that two seemingly different samples in the deformation gradient space might point towards exactly the same point in invariant space. 
Physically this can be understood by two distinct right Cauchy-Green tensors corresponding to the same principal stretches.
However, this is problematic since we do not gain any new information when using two right Cauchy-Green tensors tensors which point to the same invariants as inputs to our stress response experiment.
Two ways to overcome this issue are to: 
\begin{itemize}
    \item Sample in the space of the principal strains. 
    \item Sample directly in invariant space. 
\end{itemize}
However, when naively sampling with either one of these ideas we crucially loose any information about the training domain that we have defined in the space of the deformation gradient. To illustrate this problem, Figure \ref{fig:increasingSpredInPrincicap} shows the spread of $20,000$ points in principal invariant space sampled from differently bounded regions of the nine-dimensional deformation gradient space of eq. (\ref{eq::DefoSpace}).
Additionally, it is obvious from Figure \ref{fig:increasingSpredInPrincicap} that the points in principal space follow a specific pattern, i.e. some combinations of $I_{1}-I_{2}-I_{3}$ values are not obtainable because they do not correspond to a physical deformation gradient.
Therefore, we aim to develop an approach that evenly spreads points in the invariant space  corresponding to a certain deformation gradient training domain following eq. (\ref{eq::DefoSpace}), while maintaining that every generated sample corresponds to a physically attainable deformation. \\

In this paper we are the first to present a sampling strategy that allows generating evenly distributed physical samples in a pre-determined region of the invariant space. Consequently, 
any trained surrogate model will be more proficient by requiring less data. 
The proposed approach is based on simulated annealing \citep{van1987simulated}. 
Furthermore, the presented approach generates evenly spread samples in isotropic invariant space. Keeping these samples fixed we present a second level of sampling in the space of pseudo-invariants. This way it is possible to efficiently build upon the sample points for isotropic materials when needing to extend to anisotropic materials, and the framework allows efficient storing of sampled points in databases.  In the following we will first restrict ourselves to generating samples for the principal invariants $I_{1} - I_{2} -I_{3}$.\\

The first requirement for the sampling approach is that a bounded region in the principal invariant space should be known based on the required deformation gradient input domain. As there are different ways to achieve this, in this work we propose to generate sufficiently many samples in the deformation gradient space, i.e. here $100,000$, such that it can be assumed that the resulting points in invariant space adequately span the bounded domain. Our approximation is that the convex hull of all of these points is equivalent to the convex hull that we aim to sample in. To generate the convex hull from a set of points in three-dimensions we refer to \cite{chazelle1993optimal}. The steps undertaken to generate the convex hull are summarized in Algorithm \ref{alg::ConvexHull}.
\begin{algorithm2e}[H]
\SetAlgoLined
\KwResult{$convI$ : Convex hull of permissible points in invariant space }
\SetKwInOut{Input}{Input}
\BlankLine
\Input{ Deformation gradient component bounds $F^{L}_{ij}$ and $F^{U}_{ij}$,
Number of random samples $n$}
\BlankLine
 Sample $\bm{F}$ $n$-times \;
 \For{$i=1:n$}{
 Obtain $J = \det \bm{F}_{i}$\;
 \If{$J>0$}{
 $\bm{C}_{i} = \bm{F}_{i}^{T}\bm{F}_{i}$\;
 Obtain $I_{1}, I_{2}, I_{3}$ from $\bm{C}_{i}$\;
 Store invariants in $\bm{pI}[i,:] = [I_{1}, I_{2}, I_{3}]$;
 }
 }
\Return{$convI = \text{conv}(\bm{pI})$}
 \caption{Obtain convex hull of permissible invariant space points based on deformation gradient bounds.}\label{alg::ConvexHull}
\end{algorithm2e}
A convex hull built this way is shown in Figure \ref{fig:exampleWhyNeeded}. From this visualization it can be noted that even though the convex hull is an enclosing envelope of all the sample points, it also encloses a part of the domain that does not correspond to physical deformations, e.g. the upper part inside the convex hull that has no samples in it. Hence, even when generating new points that lie inside the convex envelope we still need to ensure that these points are physical.
A simple but often overlooked check for the physicality of an invariant set in isotropic materials is derived
from the fact that all principal strains ($\lambda_{1}, \lambda_{2}, \lambda_{3}$) defined as the square root of the positive eigenvalues of the right Cauchy-Green tensor have to be real-valued.
According to 
\cite{currie2004attainable} and \cite{burnside1892theory} and under 
consideration that three principal invariants $(I_{1},I_{2},I_{3})$ are given we can define
\begin{equation}
    \begin{aligned}
    H&= \frac{1}{9} (I_{1}^{2}- 3 I_{2}) \\
    G &= \frac{1}{3} I_{1} I_{2} - I_{3} - \frac{2}{27} I_{1}^{3} \\
    \beta &= \arccos ( - \frac{G}{2 H^{\frac{3}{2}}})
    \end{aligned}
\end{equation}
Following the theorem from \cite{burnside1892theory}[p.84, "Criterion of the Nature of the Roots of a Cubic."] we can say that when the two conditions
\begin{equation}\label{eq::PhysicalCheckIso}
    \begin{aligned}
        1. \qquad &G^{2} + 4 H^{3} \leq 0  \\  2. \qquad &\beta \in \mathbb{R} 
    \end{aligned}
\end{equation}
are fulfilled,
the triplet of invariants $(I_{1},I_{2},I_{3})$ correspond to a physical deformation.
When these conditions are met, the squares of the principal strains can be reconstructed as
\begin{equation}\label{eq::ObtainC1}
    \begin{aligned}
        \lambda_{1,rec}^{2} &=\frac{1}{3} I_{1} -  2 \sqrt{H} \cos (\frac{\pi - \beta}{3}) \\
            \lambda_{2,rec}^{2} &= \frac{1}{3} I_{1}- 2 \sqrt{H} \cos (\frac{\pi + \beta}{3})\\
    \lambda_{3,rec}^{2} &= \frac{1}{3} I_{1} + 2 \sqrt{H} \cos (\frac{\beta}{3}). 
    \end{aligned}
\end{equation}
This, crucially (for the determination of the resulting stresses) allows us to reconstructed a right Cauchy-Green tensor that corresponds to $(I_{1},I_{2},I_{3})$ 
\begin{equation}\label{eq::ObtainC2}
    \bm{C}_{rec} =\begin{bmatrix}
     \lambda_{1,rec}^{2} & 0 & 0 \\
     0 &  \lambda_{2,rec}^{2} & 0 \\
     0 & 0 &  \lambda_{3,rec}^{2}
    \end{bmatrix}.
\end{equation}
Now that we have defined an approach to evaluate the physicality of the points in the principal invariant space and are also able to ensure that the points are inside the input domain in deformation gradient space,  we can outline the simulated annealing algorithm. \\

Let us assume that our goal is to generate $N$ evenly sampled points in the invariant space.
The algorithm starts with $N-1$ randomly sampled points that are grouped in an array $\bm{pI} \in \mathbb{R}^{(N-1) \times 3}$ inside the bounded invariant space, which could for example be coming from $N-1$ sampled points in the deformation gradient space. When those initial $N-1$ points are obtained, extend $\bm{pI}$ with one more sample which has the invariant values of the undeformed configuration ($I_{1} = 3$, $I_{2} = 3$, $I_{3}=1$). This last sample is kept unchanged throughout the simulated annealing process in order to ensure the preservation of the stress-free undeformed configuration. \\

The basic idea of the algorithm is then to iteratively adjust the positions of the generated samples over $N_{T}$ annealing steps based on three checks:
\begin{enumerate}
    \item The new position of the current sample decreases the distance to the nearest neighbor of the remaining dataset in comparison to its previous position.
    \item The new sample position of the current point is inside the convex hull of the predefined bounded deformation gradient domain.
    \item The new sample position of the current sample corresponds to a physical point in principal invariant space.
\end{enumerate}

\begin{figure}[h!]
    \centering
    \includegraphics[scale=0.5]{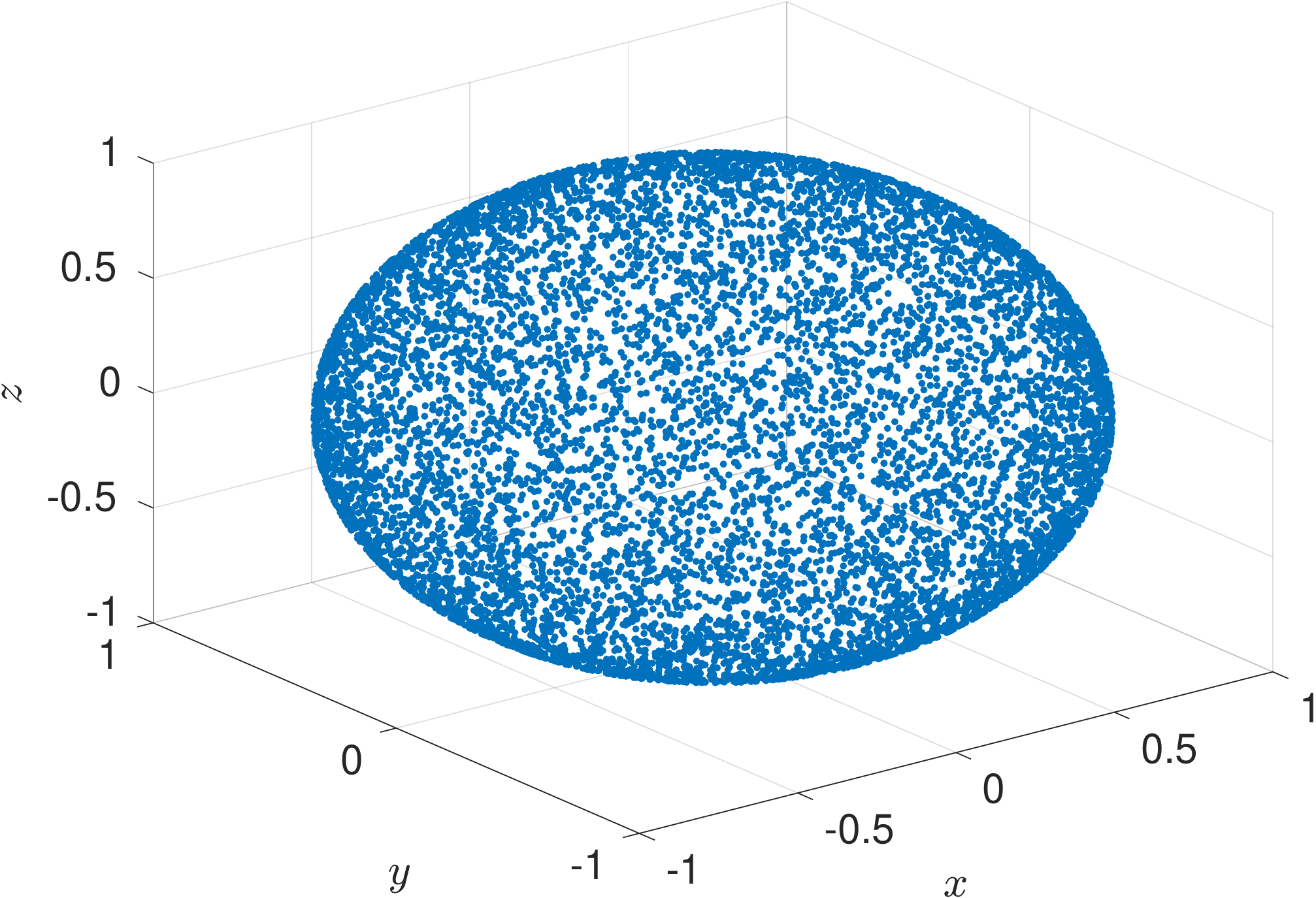}
    \caption{$10,000$ random points on the three-dimensional unit sphere generated with Box-Muller transform. }
    \label{fig:boxMuller}
\end{figure}

To do this in each annealing step we loop over every point of the current sample set $\bm{pI}$. For the current sample $\bm{pI}_{j}$, we find the distance $d$ to the closest data point in the remaining sample set $\lbrace \bm{pI} \setminus \bm{pI}_{j} \rbrace$. 
After that, we utilize the Box-Muller transform \citep{box1958note} in order to generate a random three dimensional step direction value $\bm{n}$, which is located on the surface of a three dimensional unit sphere.
The Box-Muller transform is defined by the following steps
\begin{equation}\label{eq::BoxMuller}
    \begin{aligned}
 \text{Step }1: \qquad   &u,v,w \in \mathcal{N}(0,1) \\
  \text{Step }2: \qquad   &   s = \sqrt{u^{2} + v^{2} + w^{2}} \\
  \text{Step }3: \qquad   &   \bm{n}  = \frac{1}{s} (u,v,w).
    \end{aligned}
\end{equation}
To prove the reliability of the algorithm $10,000$ generated step direction are plotted in Figure \ref{fig:boxMuller}.
\\

\begin{figure}[ht]
\begin{subfigure}[b]{0.5\linewidth}
\centering
\includegraphics[scale=0.28]{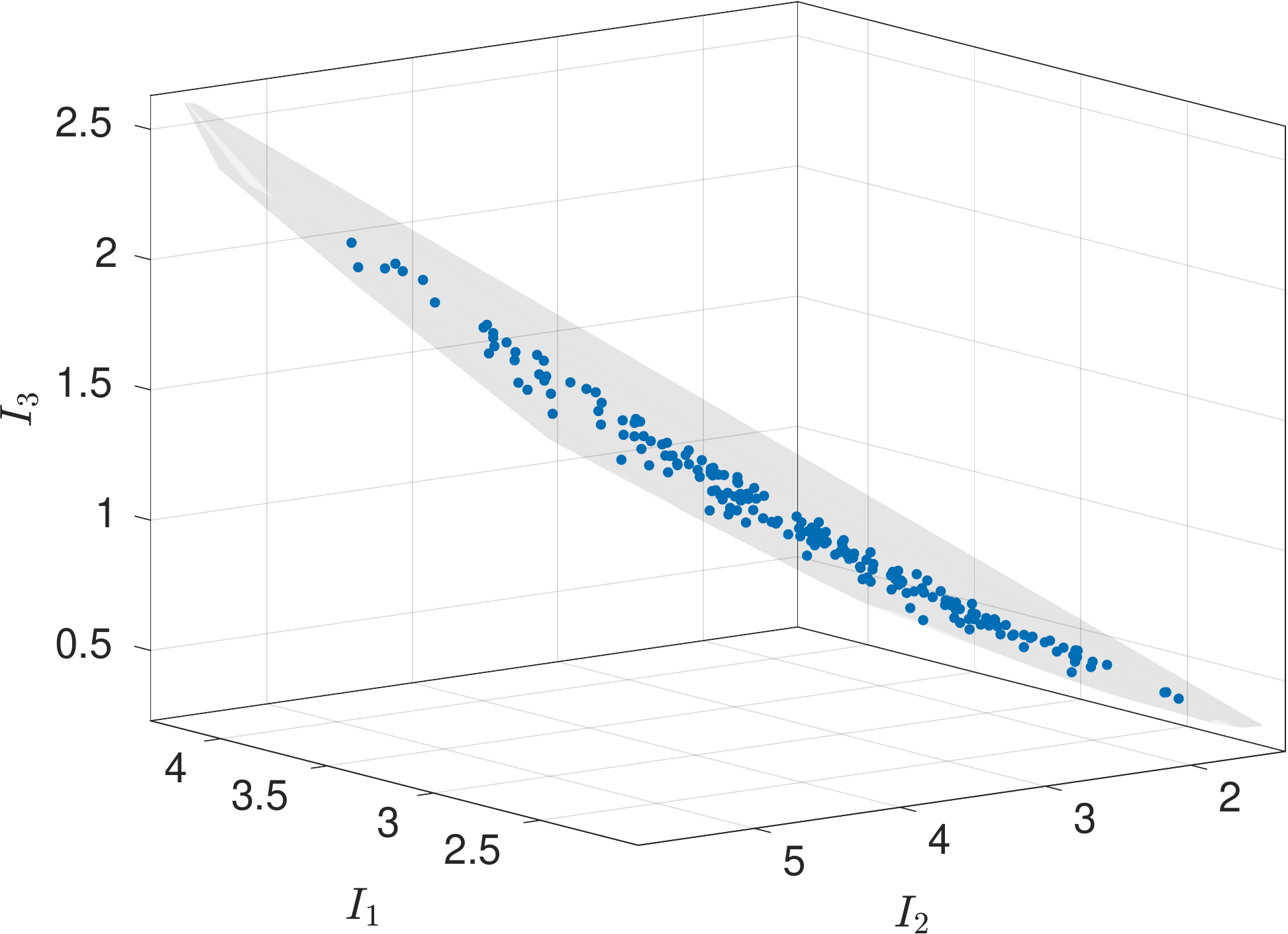} 
\caption{$n=0$}\label{fig:}
\end{subfigure}%
\begin{subfigure}[b]{.5\linewidth}
\centering
\includegraphics[scale=0.28]{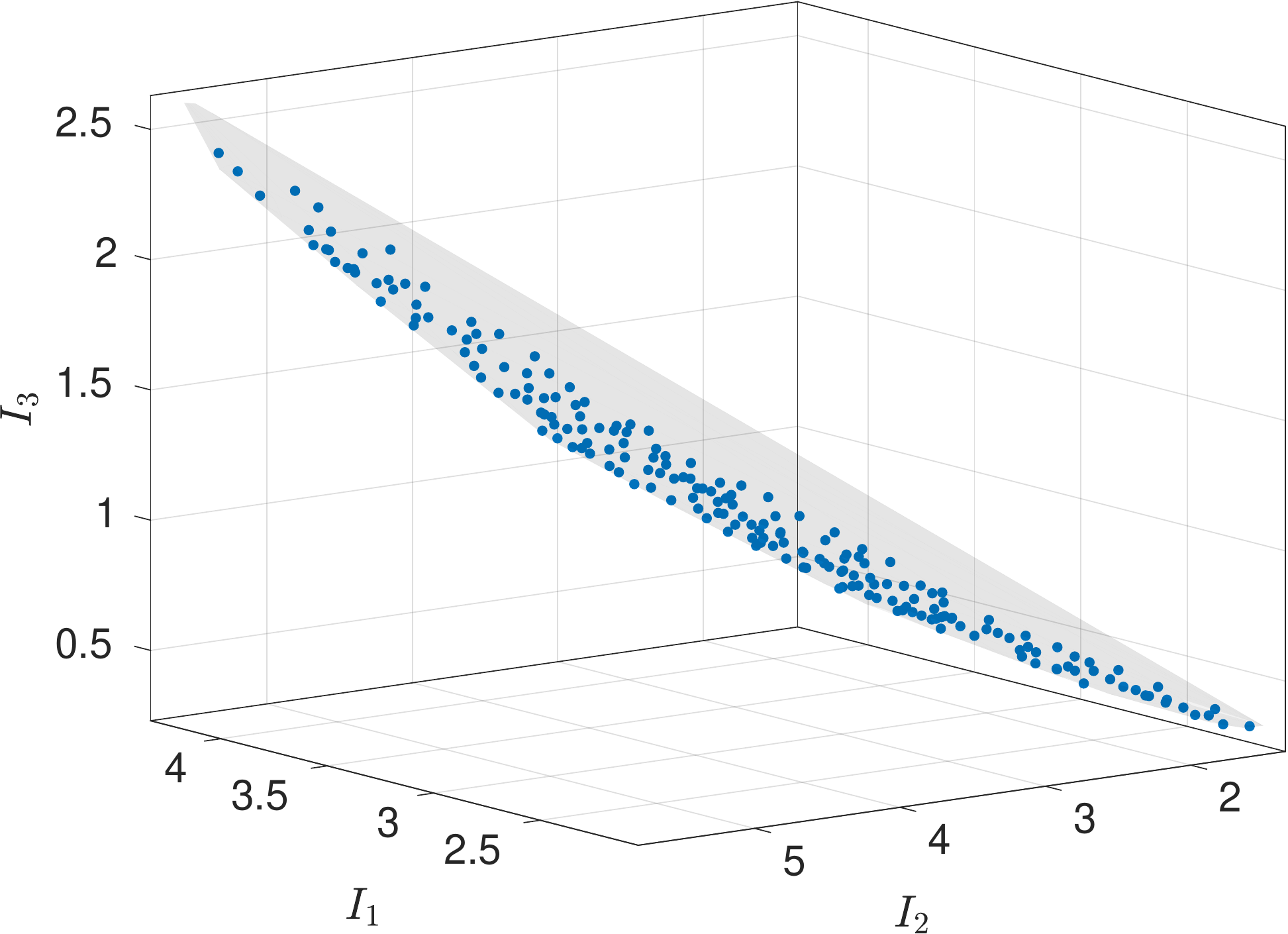} 
\caption{$n=2500$}\label{fig:}
\end{subfigure}

\begin{subfigure}[b]{0.5\linewidth}
\centering
\includegraphics[scale=0.28]{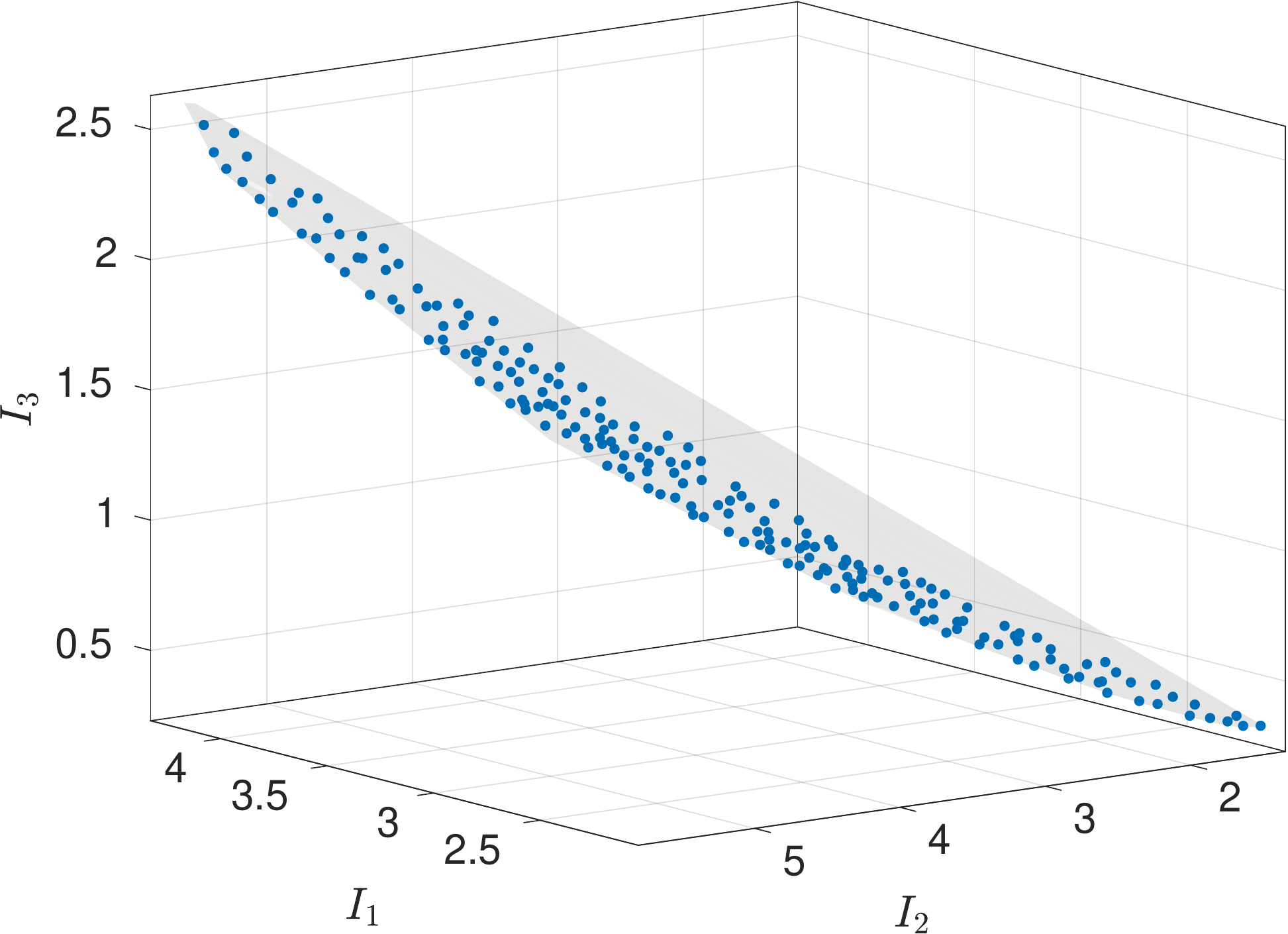} 
\caption{$n=5000$}\label{fig:IsoSteps}
\end{subfigure}%
\begin{subfigure}[b]{.5\linewidth}
\centering
\includegraphics[scale=0.28]{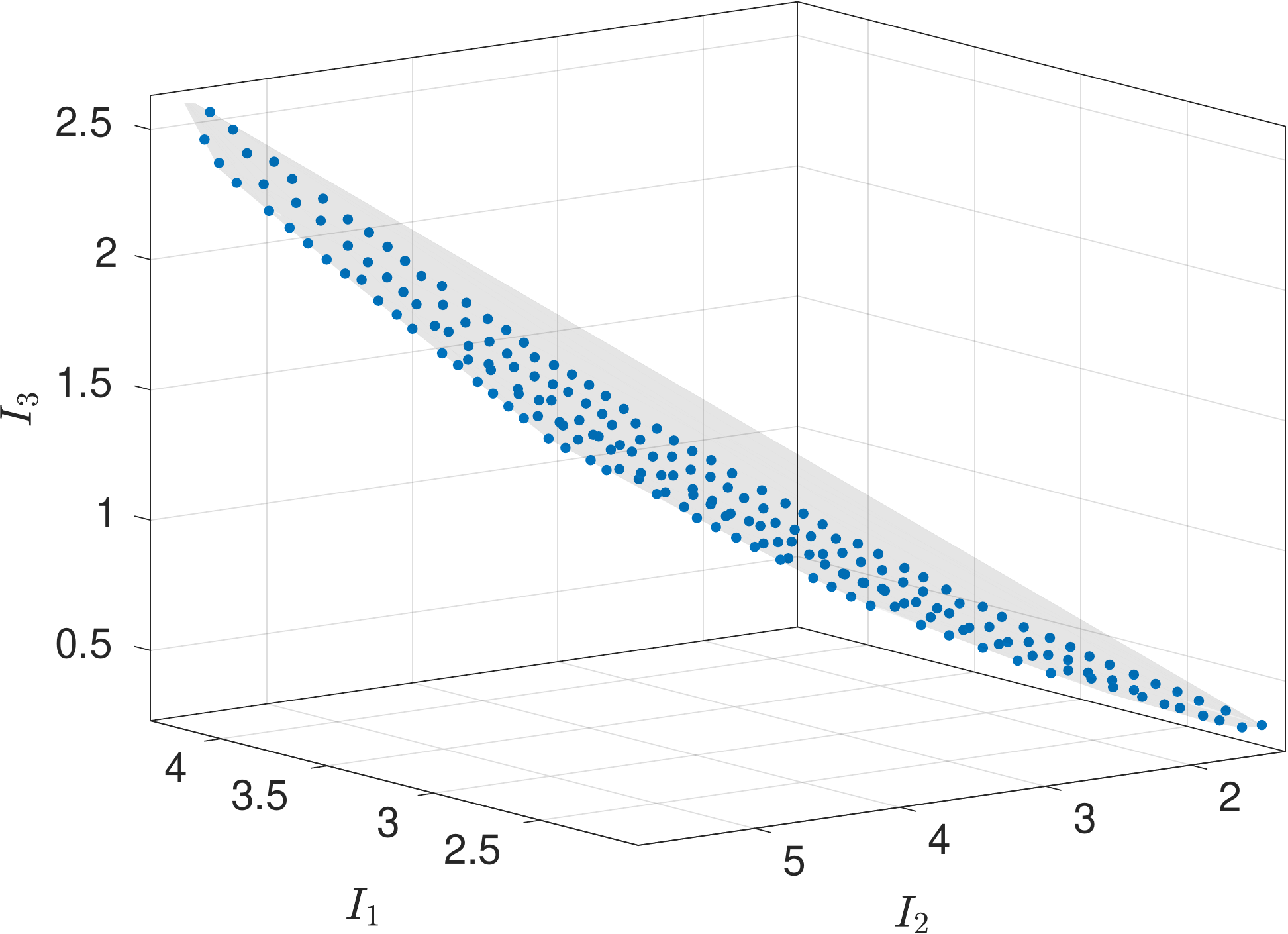} 
\caption{$n=7500$}\label{fig:}
\end{subfigure}
\caption{Selected steps of the proposed minimax (space-filling) distance sampling approach with $200$ sample points in invariant space $(I_{1}-I_{2}-I_{3})$. Convex hull of permissible sample point positions of $17.5\%$ input range in deformation gradient space in grey.
Step size $T=1$,
Step size factor $\alpha= 0.9995$.}\label{fig:}
\end{figure}

Next, we sample a step magnitude $s\in [0,T]$ where $T$ is the current step size. Now that we know the step size and step direction we can obtain the next possible position of the sample $\bm{pI}_{j}$ with $\bm{p} = \bm{pI}_{j} + s \bm{n}$.
This point position is accepted ($\bm{pI}_{j} = \bm{p}$) when i) its distance to the nearest neighbor in the remaining dataset is larger than $d$, when ii) it passes the physicality test of eq. (\ref{eq::PhysicalCheckIso}), and iii) it is inside the convex hull of the pre-sampled points of algorithm \ref{alg::ConvexHull}. If one of these conditions is not met,  $\bm{pI}_{j}$ remains unchanged. This process is repeated $N_{T}$ times for every single point in the dataset. After each iteration the step size $T$ is reduced by a constant factor $T = \alpha T$ where $0< \alpha < 1$. 
The full algorithm is summarized in Algorithm box \ref{algo::EvenIso}.
\begin{algorithm2e}[H]
\SetAlgoLined
\KwResult{$\bm{pI}$ : Matrix of evenly spread points in invariant space }
\SetKwInOut{Input}{Input}
\BlankLine
\Input{ Convex hull of permissible points $convI$,
 Target number of points in invariant space $N$,
Number of annealing steps  $N_{T}=7,000$,
Step size $T=1$,
Step size factor $\alpha= 0.9995$}
\BlankLine
Sample $N-1$ points randomly in $convI$ generating $\bm{pI}$\;
Add vector of unstressed configuration $I_{1} = 3$, $I_{2} = 3$, $I_{3}=1$ to $\bm{pI}$\;
 \For{$i=1:N_{T}$}{
 \For{$j=1:N$}{
 Set $\bm{pN} = \lbrace \bm{pI}\setminus \bm{pI}_{j}\rbrace$  \;
 Set $d$ as distance to closest neighbor of $\bm{pI}_{j}$ in $\bm{pN}$ \;
 Randomly sample $3$-dimensional unit sphere point $\bm{n}$  (eq. (\ref{eq::BoxMuller}))\;
 Set  $s \in \mathcal{U}[0,T]$\;
 $\bm{p} = \bm{pI}_{j} + s \bm{n}$ \;
  Set $d_{test}$ as distance to closest neighbor of $\bm{p}$ in $\bm{pN}$ \;
 $t= 0$, $t_{u}=1$\;
  \If{$\bm{pI}_{j} == [3,3,1] $}{$t_{u}=0$}
 \If{$\bm{p}$ in $convI$ \&\& $d_{test} >d$  \&\& $\bm{p}$ is physical (\text{see eq. (\ref{eq::PhysicalCheckIso}}))}{$t=1$}

 $\bm{pI}_{j} = \bm{pI}_{j} + s \,t \, t_{u} \,\bm{n} $ 
 }
 $T = \alpha T$ 
 }
\Return{$\bm{pI}$}
 \caption{Algorithm to obtain evenly spread points in invariant space based on physical constraints and deformation gradient bounds using a simulated annealing approach. Values chosen by the authors are provided.}\label{algo::EvenIso}
\end{algorithm2e}
The effectiveness of the proposed approach is highlighted by an example. Consider a $17.5\%$ training domain in which we aim to generate $200$ samples. We use an initial step size of $T=1$, and a step size factor of $\alpha= 0.9995$.
The sample positions over the process after a different number of steps have been completed are shown in Figure \ref{fig:IsoSteps} along with the convex hull of the bounded domain in grey. It can be seen that the initially generated samples are not spread evenly in the intended region. However after $7,500$ steps a space-filling sample distribution can be observed. 

\subsection{Extension to anisotropic materials}
\begin{figure}
    \centering
    \includegraphics[scale=0.4]{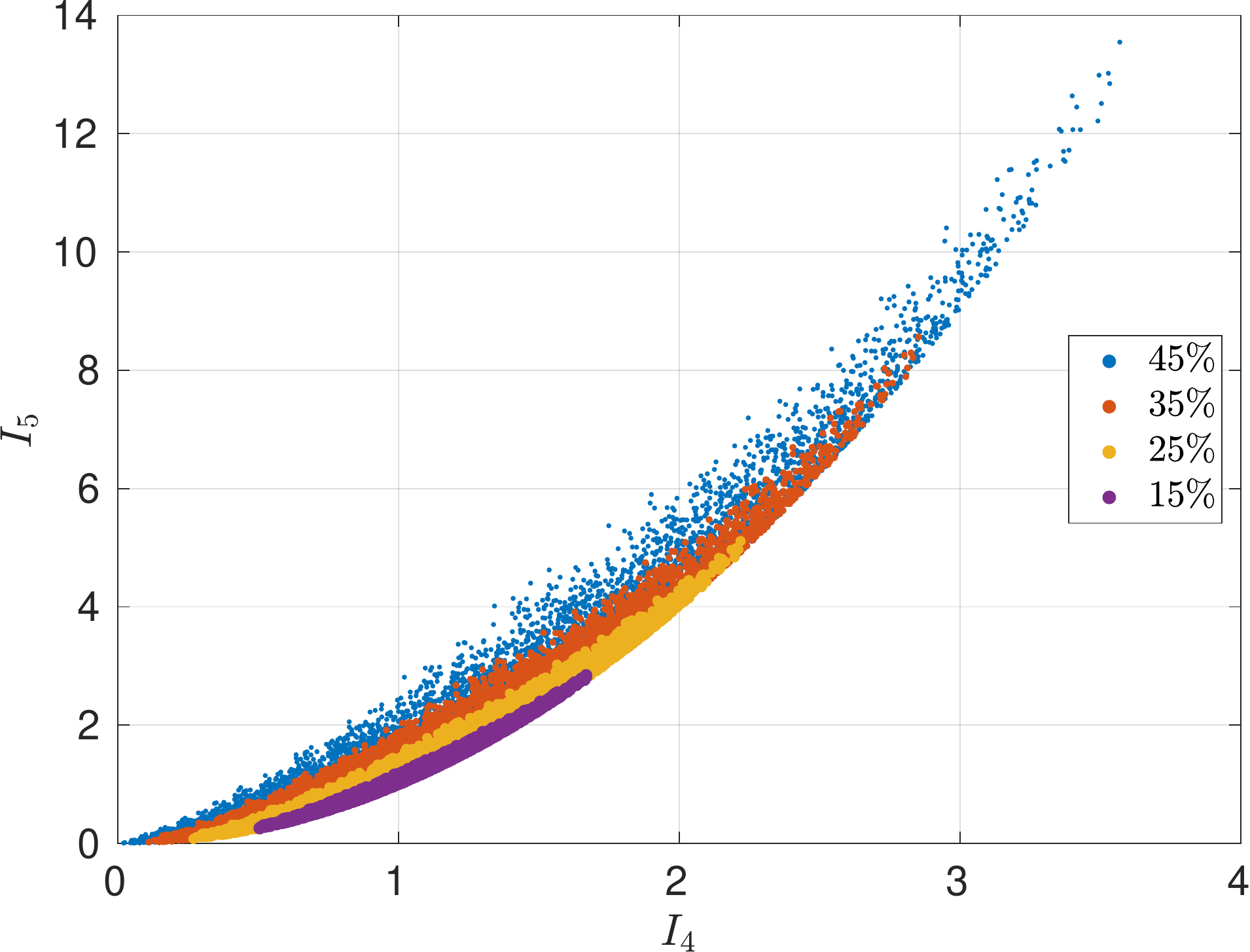}
    \caption{Spread of pseudo-invariants with increasing training domain and $\bm{a}_{0}= \frac{1}{\sqrt{6}}[1,2,1]^{T}$. $10,000$ samples are plotted in deformation gradient space with LHS in each bounded domain. Last two components of the quintuple ($I_{1},I_{2},I_{3},I_{4}, I_{5}$) are plotted.}
    \label{fig:spreadOfPseudo}
\end{figure}

\begin{figure}[ht]
\begin{subfigure}[b]{0.5\linewidth}
\centering
\includegraphics[scale=0.35]{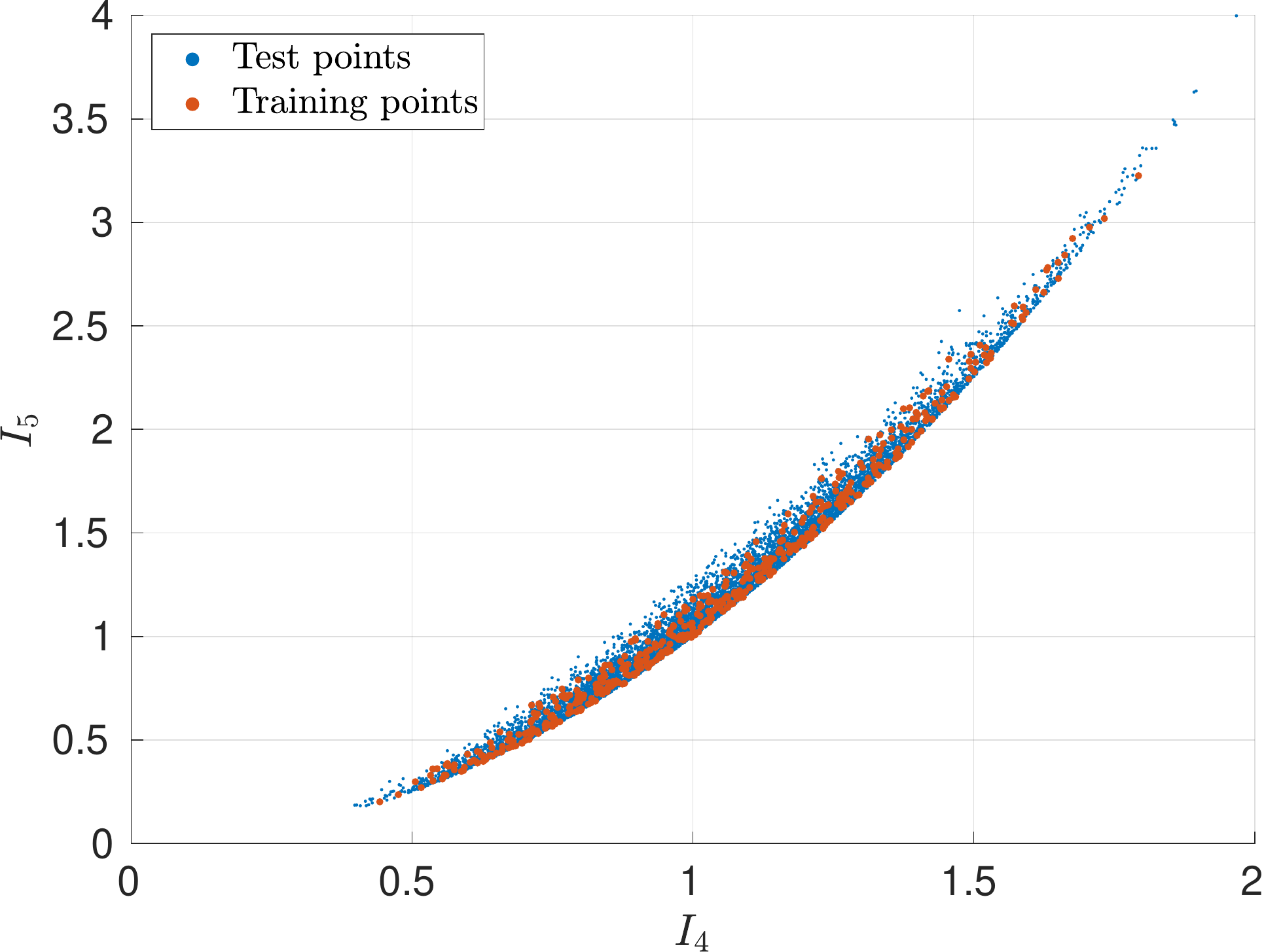} 
\caption{Initial sample positions projected onto $I_{4}-I_{5}$.}\label{fig::SamplingI4I5a}
\end{subfigure}%
\begin{subfigure}[b]{0.5\linewidth}
\centering
\includegraphics[scale=0.35]{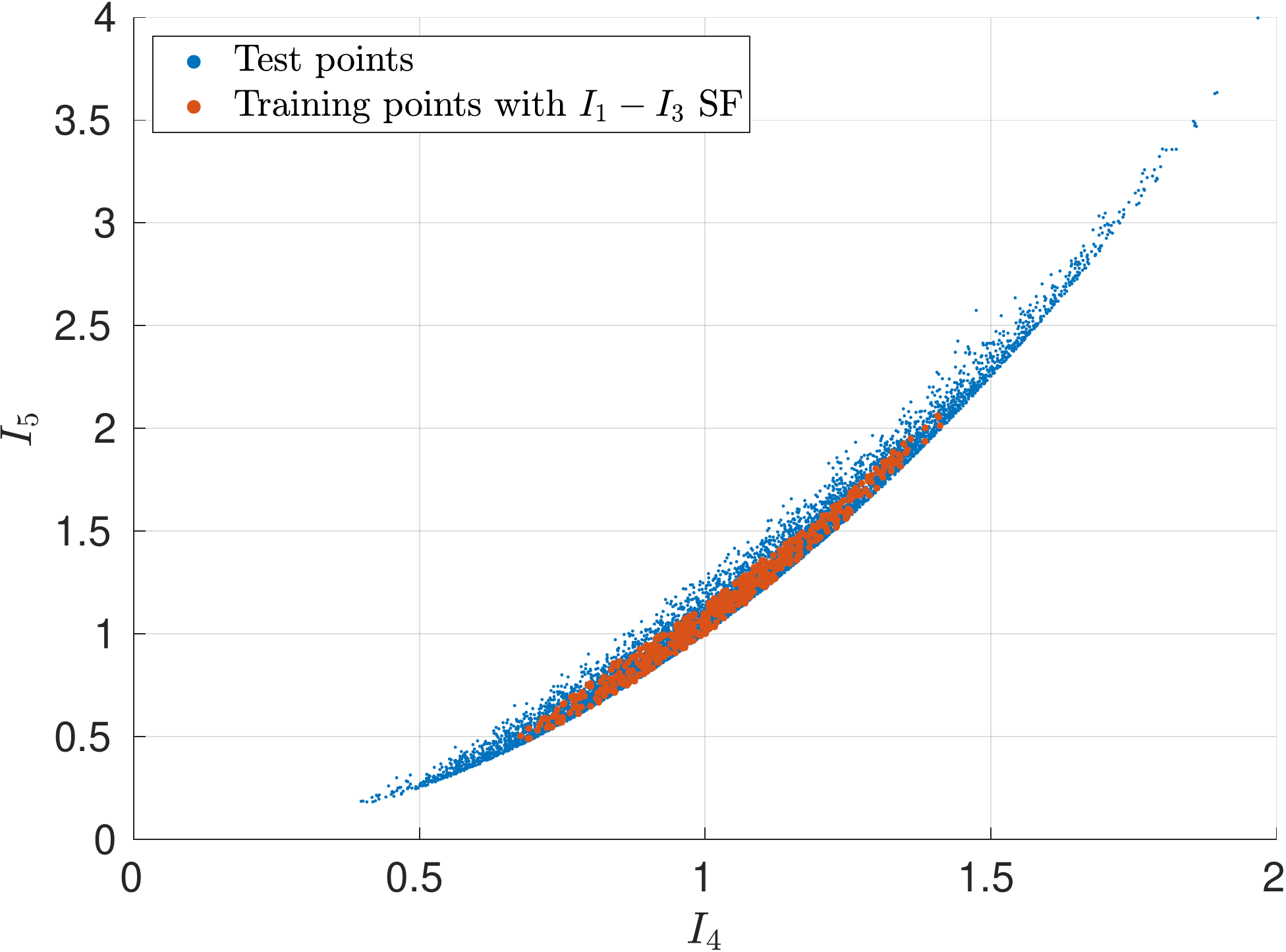} 
\caption{Initial sample positions projected onto $I_{4}-I_{5}$.}\label{fig::SamplingI4I5b}
\end{subfigure}
\begin{subfigure}[b]{1.0\linewidth}
\centering
\includegraphics[scale=0.35]{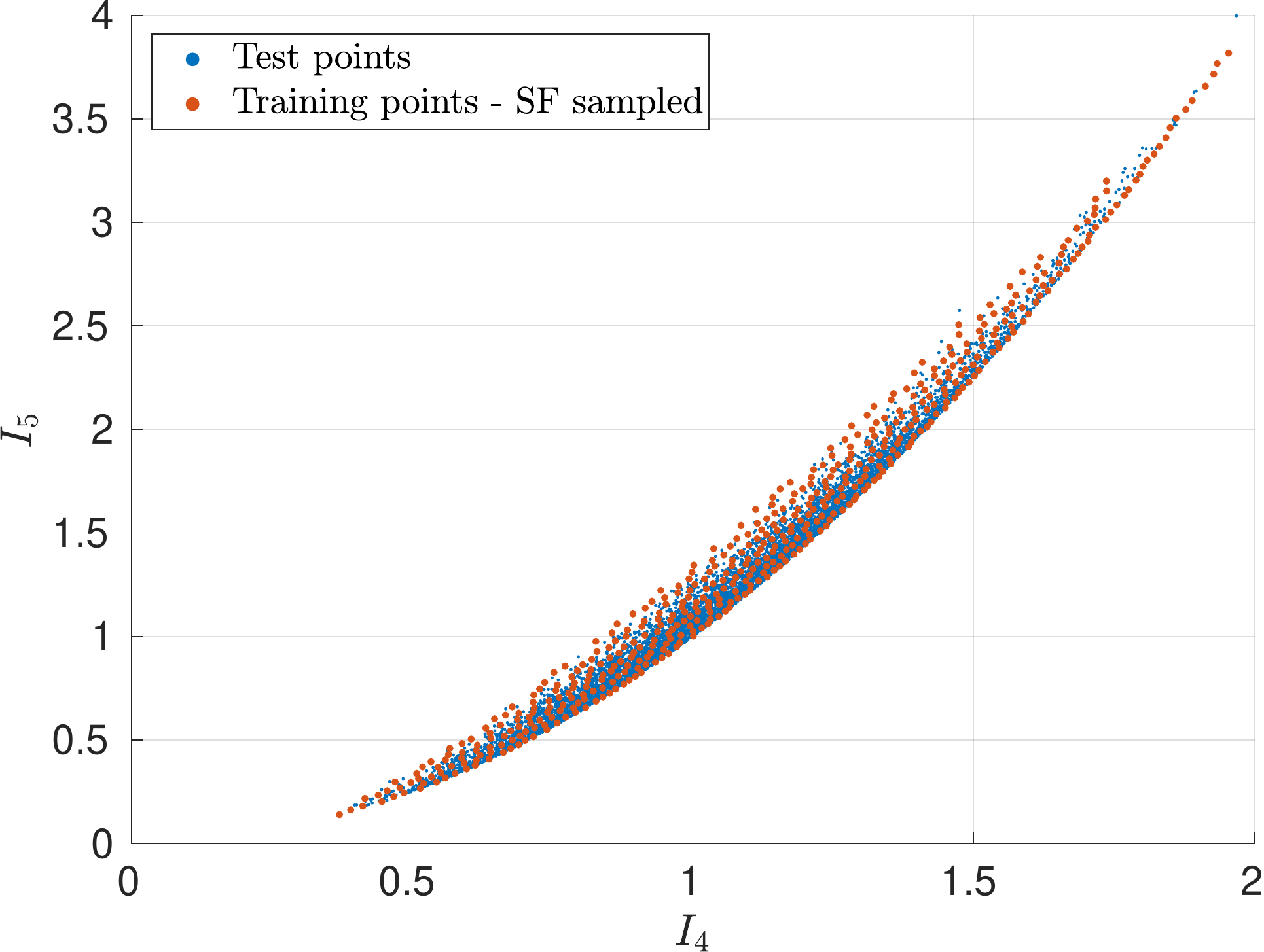} 
\caption{Sample positions after $10,000$ annealing steps of proposed space-filling approach projected onto $I_{4}-I_{5}$.}\label{fig::SamplingI4I5c}
\end{subfigure}
\caption{Initial positions and positions after minimax approach with $10,000$ iteration steps of $500$ samples projected onto $I_{4}-I_{5}$ plane in red ($\bm{a}_{0}= \frac{1}{\sqrt{6}}[1,2,1]^{T}$). Positions of $10,000$ test points generated in $17.5\%$ training range in deformation gradient space with LHS and then projected onto $I_{4}-I_{5}$ plane in blue.}\label{fig::SamplingI4I5}
\end{figure}

In the previous section we have discussed how to generate evenly spaced physical samples inside a bounded domain. However, the presented algorithm only applied to the principal invariants ($I_{1},I_{2},I_{3}$) and is not applicable to finding equivalent space-filling components of the pseudo invariant space that includes pseudo-invariants (e.g. in the transversally isotropic case $I_{4}, I_{5}$). This is due to the fact that there is, to the best of the authors knowledge, no direct analytical check for the physicality/reachability of the quintuple ($I_{1},I_{2},I_{3},I_{4}, I_{5}$), i.e. if a specific combination of the five invariants is realizable with real eigenvalues of the Cauchy-Green tensor where ($I_{1}, I_{2}, I_{3}$) are inside a bounded domain.
In Figure \ref{fig:spreadOfPseudo} we sampled $10,000$ samples in deformation gradient space for different training domains (corresponding to increasing values of $\delta\in[0.15,0.45]$ from eq. (\ref{eq::DefoSpace})) and plotted the transversally isotropic pseudo-invariants (assuming a known referential vector $\bm{a}_{0}= \frac{1}{\sqrt{6}}[1,2,1]^{T}$) for the generated samples. We can see that all of the pseudo-invariant values follow a clear pattern (not all points are reachable, i.e. the values $I_{4}=I_{5}=2$) and furthermore that the convex-hull of the points would not be a good enough indicator as to where pseudo-invariant points can reside.
Hence we need to come up with an additional sampling algorithm for the pseudo-invariant space, which is independent of the spread of the primary invariants. 
That the  samples in the pseudo-invariant space must also be evenly distributed can be seen from Figures \ref{fig::SamplingI4I5a} and \ref{fig::SamplingI4I5b}. In Figure \ref{fig::SamplingI4I5} we again sampled $10,000$ points from the deformation gradient space with LHS in a $17.5 \%$ training domain (blue dots) as well as an additional 500 separate samples (red) dots. It can be seen that especially the outer edges of the envelope surrounding the blue dots are not well preserved by the $500$ samples.\\

Next, we used algorithm \ref{algo::EvenIso} to sample $500$ evenly spaced points in the primary invariant space. For each of these triplet of ($I_{1}, I_{2}, I_{3}$) we used eqs. (\ref{eq::ObtainC1}) and (\ref{eq::ObtainC2}) to a obtain corresponding right Cauchy-Green tensor. Since we know the referential vector $\bm{a}_{0}$ we are able to obtain the pseudo-invariant values ($I_{4}, I_{5}$) for all of the $500$ samples which are plotted in Figure \ref{fig::SamplingI4I5b}. It can be seen that even though we have an excellent spread in the primary invariant space the obtained samples in pseudo-invariant space are very clustered and are not at all space-filling.\\

Hence, we propose an additional algorithm based on simulated annealing that is able to generate evenly spaced points in the pseudo-invariant space. The final points generated by this algorithm (using the same settings as described for Figures \ref{fig::SamplingI4I5a} and \ref{fig::SamplingI4I5b}) are shown in Figure \ref{fig::SamplingI4I5c} which preemptively highlights the performances of the following procedure.  
The idea behind this approach, in the context of anisotropic material response, is based upon the fact that simple rotations of the right Cauchy Green tensor result in different pseudo-invariant values while leaving the principal invariants unchanged. To emphasize this we look at the following example. Consider a right Cauchy-Green tensor to be given by 
\begin{equation}
    \bm{C} = \begin{bmatrix}
    1.4 &	0 &0 \\
0 &	1.1 &	0 \\
0 &0 & 0.8
    \end{bmatrix}
\end{equation}
which using $\bm{a}_{0}= \frac{1}{\sqrt{6}}[1,2,1]^{T}$ has the invariant quintuple ($3.3, 3.54, 1.232, 1.1, 1.24$). Next consider this tensor to be rotated in $y-z$ plane around an angle of $0.1$ rad, i.e.
\begin{equation}
    \bm{C}_{rot} = \bm{R}_{\alpha=0.1}^{T} \bm{C} \bm{R}_{\alpha=0.1} \approx
    \begin{bmatrix}
    1.4 &	0	&0 \\
0 &1.097&  -0.029\\
0 & -0.029 & 0.803
    \end{bmatrix}
\end{equation}
which has the approximate invariant quintuple ($3.3,3.54,1.232,1.078,1.199$). Hence, this observation allows us to generate different pseudo-invariant values from known (and fixed) right Cauchy-Green tensors. We take advantage of this and build a simulated annealing algorithm around the three angles defining the rotation matrices acting on the three planes $x-y$, $x-z$ and $y-z$ in euclidean space.  This has the advantage that the generated quintuple of values ($I_{1},I_{2},I_{3},I_{4},I_{5}$) is always reachable from the primary invariants and the already obtained evenly spaced primary invariants do not need to be changed.
This algorithm starts with the knowledge about $N$ evenly spaced principal invariants $\bm{pI}_{Iso} \in \mathbb{R}^{N \times 3}$ in some confined space with the requirement that one of the components of $\bm{pI}$ has to correspond to the invariants of the undeformed configuration. Then, using eqs. (\ref{eq::ObtainC1}) and (\ref{eq::ObtainC2}) we obtain the right Cauchy-Green tensor $\bm{C}_{j}$, $j=1,\ldots,N$ in principal space for each of the $N$ samples and the corresponding pseudo-invariants which are stored in $\bm{pI}_{ani} \in \mathbb{R}^{N \times 2}$. For simplicity we consider the three rotation matrices 
\begin{equation}
    \bm{R}_{x} = \begin{bmatrix}
\cos{\alpha_{x}} & -\sin{\alpha_{x}} & 0 \\
\sin{\alpha_{x}} & \cos{\alpha_{x}} & 0 \\
0 & 0 & 1
\end{bmatrix}, \qquad \bm{R}_{y} = \begin{bmatrix}
\cos{\alpha_{y}} & 0 & \sin{\alpha_{y}} \\
0 & 1 & 0 \\
-\sin{\alpha_{y}} & 0 & \cos{\alpha_{y}}
\end{bmatrix}, \qquad \bm{R}_{z} = \begin{bmatrix}
1 & 0 & 0 \\
0 & \cos{\alpha_{z}} & -\sin{\alpha_{z}} \\
0 & \sin{\alpha_{z}} & \cos{\alpha_{z}}
\end{bmatrix}.
\end{equation}
Each component of $\bm{pI}_{ani}$ gets assigned to one instance of the three rotation defining angles which are initially individually sampled in $\mathcal{U}[0, 2\pi]$ and stored in the angle matrix $\bm{\mathcal{A}} \in \mathbb{R}^{N \times 3}$. \\

Then, similarly to Algorithm \ref{algo::EvenIso} the process iterates over $N_{T}$ loops where each loop consists of the following procedure. Loop over all elements of $\bm{pI}_{ani}$.
For each component $\bm{pI}_{j,ani} \subset \bm{pI}_{ani}$ find the distance $d$ to the current closest neighbor in the remaining dataset. Then using Box-Muller transform of eq. (\ref{eq::BoxMuller}) sample a random point on the three dimensional unit sphere which acts as the step direction $\bm{n}$ of the current sample. Then, knowing the current step size $T$, sample $s=\mathcal{U}[0,T]$. The next possible position of the angle $\bm{\mathcal{A}}_{j}$ that is assigned to the current point can then be obtained with $\bm{p} =\bm{\mathcal{A}}_{j} + s \bm{n}$. In order to check if this set of angles leads to a pseudo-invariant combination that is further away from the closest point than $d$, we obtain the rotated right Cauchy-Green tensor of the current point with
\begin{equation}
    \bm{C}_{test} = \bm{R}_{x}^{T}(\bm{p}[1]) \bm{R}_{y}^{T}(\bm{p}[2]) \bm{R}_{z}^{T}(\bm{p}[3]) \bm{C}_{j} \bm{R}_{z}(\bm{p}[3])  \bm{R}_{y}(\bm{p}[2]) \bm{R}_{x}(\bm{p}[1]).
\end{equation}
From this tensor we can again obtain $I_{4,test}$ and $I_{5,test}$ and use these values to check the distance to the closest neighbor in the remaining dataset $\lbrace \bm{pI}_{ani}\setminus \bm{pI}_{j,ani} \rbrace$. If this distance is larger than $d$, set $\bm{\mathcal{A}}_{j} =\bm{\mathcal{A}}_{j} + s \bm{n} $ and $\bm{pI}_{j,ani} =[I_{4,test}, I_{5,test}]$. If this is not the case, then leave all values unchanged.
After each loop over all points of $\bm{pI}_{ani}$ reduce the step size with $T=\alpha T$ where $0<\alpha<1$.
The full approach is summarized in algorithm box \ref{algo:PseudoSpread}.

\begin{algorithm2e}[h!]
\SetAlgoLined
\KwResult{$\bm{pI} \in \mathbb{R}^{N \times 5}$ : Matrix of evenly spread points in isotropic-invariant ($I_{1}, I_{2}, I_{3}$) and pseudo-invariant ($I_{4}, I_{5}$) space }
\SetKwInOut{Input}{Input}
\BlankLine
\Input{ 
 Target number of points in invariant space $N$,
Number of annealing steps  $N_{T}=10,000$,
Step size $T=2 \pi$,
Step size factor $\alpha=0.9995$, Evenly sampled points in isotropic invariant space $\bm{pI}_{iso} \in \mathbb{R}^{N \times 3}$ based on Algorithm \ref{algo::EvenIso}}
\BlankLine
\For{$j=1:N$}{
Obtain $\bm{C}_{j}$ from $\bm{pI}_{j, iso}$ using equations (\ref{eq::ObtainC1}) and (\ref{eq::ObtainC2})\;
Set $\bm{pI}_{j,ani}=[I_{4}, I_{5}]$ obtained from $\bm{C}_{j}$ 
}
Initialize angle matrix $\bm{\mathcal{A}} \in \mathbb{R}^{N \times 3} \in \mathcal{U}[0, 2 \pi]$ \;
Set $\bm{R}_{x}(\alpha_{x}) = \begin{bmatrix}
\cos{\alpha_{x}} & -\sin{\alpha_{x}} & 0 \\
\sin{\alpha_{x}} & \cos{\alpha_{x}} & 0 \\
0 & 0 & 1
\end{bmatrix}$, $\bm{R}_{y}(\alpha_{y}) = \begin{bmatrix}
\cos{\alpha_{y}} & 0 & \sin{\alpha_{y}} \\
0 & 1 & 0 \\
-\sin{\alpha_{y}} & 0 & \cos{\alpha_{y}}
\end{bmatrix}$, $\bm{R}_{z}(\alpha_{z}) = \begin{bmatrix}
1 & 0 & 0 \\
0 & \cos{\alpha_{z}} & -\sin{\alpha_{z}} \\
0 & \sin{\alpha_{z}} & \cos{\alpha_{z}}
\end{bmatrix}$\;
 \For{$i=1:N_{T}$}{
 \For{$j=1:N$}{
 Set $\bm{pN}_{ani} = \lbrace \bm{pI}_{ani}\setminus \bm{pI}_{j,ani} \rbrace$  \;
 Set $d$ as distance to closest neighbor of $\bm{pI}_{j,ani}$ in $\bm{pN}_{ani}$ \;
 Randomly sample $3$-dimensional unit sphere point $\bm{n}$  (eq. (\ref{eq::BoxMuller}))\;
 Set $s \in \mathcal{U}[0,T]$\;
 $\bm{p} = \bm{\mathcal{A}}_{j} + s \bm{n}$ \;
 Set $\bm{C}_{test} = \bm{R}_{x}^{T}(\bm{p}[1]) \bm{R}_{y}^{T}(\bm{p}[2]) \bm{R}_{z}^{T}(\bm{p}[3]) \bm{C}_{j} \bm{R}_{z}(\bm{p}[3])  \bm{R}_{y}(\bm{p}[2]) \bm{R}_{x}(\bm{p}[1])$ \;
 Set $\bm{p}_{ani}= [I_{4}, I_{5}]$ obtained from $\bm{C}_{test}$ \;
  Set $d_{test}$ as distance to closest neighbor of $\bm{p}_{ani}$ in $\bm{pN}_{ani}$ \;
\If{$d_{test} >d$  }{ $\bm{\mathcal{A}}_{j}   =\bm{\mathcal{A}}_{j} + s \bm{n} t  $ \;
$\bm{pI}_{j,ani} = \bm{p}_{ani}$
}

 }
 $T = \alpha T$ 
 }
\Return{$\bm{pI} = [\bm{pI}_{iso}, \bm{pI}_{ani}]$}
 \caption{Obtain evenly spread samples when pseudo-invariants are present. Values chosen by the authors are provided.}\label{algo:PseudoSpread}
\end{algorithm2e}

In order to obtain stress responses from some sample ($I_{1},I_{2},I_{3},I_{4},I_{5}$), a corresponding right Cauchy-Green tensor needs to be available. As previously discussed, no analytical version analogue to the one for isotropic materials (eqs. (\ref{eq::ObtainC1}) and (\ref{eq::ObtainC2})) has been presented in the literature. For this reason we propose an approach based on numerical optimization.
The right Cauchy-Green tensor is a solution of the constrained nonlinear equation system
\begin{equation}
\begin{aligned}
\text{find } &\bm{C}^{\star} \text{ such that} 
\begin{bmatrix}
\text{tr}(\bm{C}^{\star}) - I_{1} \\
0.5 (\text{tr}(\bm{C}^{\star})^{2} - \text{tr}((\bm{C}^{\star})^{2})) - I_{2} \\
\text{det}(\bm{C}^{\star})- I_{3} \\
\text{tr}(\bm{A} \bm{C}^{\star})  - I_{4}\\
\text{tr}(\bm{A} (\bm{C}^{\star})^{2}) - I_{5}
\end{bmatrix}
&= 
\begin{bmatrix}
0\\
0\\
0 \\
0\\
0
\end{bmatrix}
\text{  where }\bm{C}_{ij}^{lb} \leq \bm{C}_{ij}^{\star}\leq \bm{C}_{ij}^{ub}
\end{aligned}
\end{equation}
where the upper and lower bounds $\bm{C}^{ub}$ and $\bm{C}^{lb}$
are known from the training domain of the deformation gradient. A real solution can be obtained with any nonlinear solver (such as the fsolve function of scipy \citep{2020SciPy-NMeth}) when the input quintuple ($I_{1},I_{2},I_{3},I_{4},I_{5}$) is a reachable value which is guaranteed when obtaining the samples from Algorithm \ref{algo:PseudoSpread}.\\
Overall, the presented sampling approach is generally applicable and is not specific to certain constitutive responses. Hence, the introduced method can generate samples for a multitude of different problems.

\section{Gaussian process regression}\label{sec::2}
Gaussian process regression has recently gained more popularity for building surrogate models for constitutive laws. This is due to their convergence guarantees, deep stochastic background and excellent performance for out-of-sample model predictions \citep{rasmussen2003gaussian}.
They have also evolved into a common choice for building active learning models \citep{fuhg2020state}. \\
Consider a general dataset consisting of $N$ data points to be given by
\begin{equation}
    \mathcal{D} = \lbrace \bm{x}_{i}, \bm{y}_{i} \rbrace_{i=1}^{N}
\end{equation}
where $\bm{x} \in \bm{R}^{n_{i}}$ and $\bm{y} \in \bm{R}^{n_{o}}$.
The output vectors can be recombined to build the output vector $\bm{ y}^{tp} \in \mathbb{R}^{n_{0} N }$ such that
\begin{equation}
\bm{ y}^{tp}=\begin{bmatrix} {\bf y}_{1} & \hdots & {\bf y}_{N} \end{bmatrix}^{T}\, .
\end{equation}

 
Assume that the input output relationship can be approximated by a realization of a Gaussian process given by
\begin{equation}\label{eq:Gauss_general}
\bm{ Y}(\bm{ x}) = \bm{\mu} + \bm{ A} \bm{ Z}\, ,
\end{equation}
with the output $\bm{ Y} \in \mathbb{R}^{n_{0}}$, the mean $\bm{\mu}  \in \mathbb{R}^{n_{0}}$, a positive-definite matrix $\bm{ A} \in \mathbb{R}^{n_{0} \times n_{0}}$ (the first set of unknown parameters), and a vector of mutually independent Gaussian processes $\bm{ Z} \in \mathbb{R}^{n_{0}}$ \citep{svenson2010multiobjective}.

The correlation decay between two inputs $\bm{ x}$ and $\bm{ x}'$ is typically modeled by a user-defined autocorrelation function.
\cite{laurent2019overview} identified the class of Mat\'{e}rn kernels as the most proficient autocorrelation formulation for computer experiments when no prior knowledge is available.
In this, we restrict ourselves to the so-called  Mat\'{e}rn 3/2 function \citep{matern1960spatial} which reads
\begin{equation}\label{eq:autocorr}
\begin{aligned}
R(\bm{ x}, \bm{ x}', \bm{\theta}_{i})  =  \prod_{k=1}^{n_{i}}  \left( 1 + \dfrac{\sqrt{3} \abs{x_{k} - x'_{k}}}{\theta_{i,k}} \right) \exp \left(-\dfrac{\sqrt{3} \abs{x_{k} - x'_{k}} }{\theta_{i,k}}  \right) \, \text{,}
\end{aligned}
\end{equation}
where $\boldsymbol{\theta}=\left[ \boldsymbol{\theta}_{1}, \ldots, \boldsymbol{\theta}_{n_{o}} \right]^{T}$ is a vector of unknown and trainable parameters

Consider the correlation matrix $\bm{R} \in \mathbb{R}^{n_{o}\times n_{o}}$ to be given by
\begin{equation}
\bm{R}({\bm{x}},\bm{x}') = \text{diag}\lbrace R(\bm{ x}, \bm{ x}', \boldsymbol{\theta}_{1}), \cdots  , R(\bm{ x}, \bm{ x}', \boldsymbol{\theta}_{n_{0}}) \rbrace\, .
\end{equation}
which allows us to write the covariance between two input values as
\begin{equation}
Cov(\bm{ Y}(\bm{ x}), \bm{ Y}(\bm{ x}')) = \bm{ A} \bm{ R}(\bm{ x},\bm{ x}')\bm{ A}^{T}
\end{equation}
and the block-component-wise entries of the covariance matrix $\bm{\Sigma} \in \mathbb{R}^{n_{o}N \times n_{o} N}$ by
\begin{equation}
[\boldsymbol{\Sigma}]_{ij} = 
Cov({\bm {Y}}({\bm{ x}}_{i}), {\bm{ Y}}({\bm{ x}}_{j})).
%
\end{equation}
Then, a prediction with GPR at the input point $\bm{ x}_{\star}$ can be made with
\begin{equation}\label{eq:mean}
\begin{aligned}
\hat{\bm{y}}({\bm{ x}}_{\star}) =  \hat{\bm{\mu}} + \boldsymbol{{\Pi}}({\bm{x}}_{\star}) \boldsymbol{\Sigma} \big({\bm{ y}}^{tp}- {\bf \mathcal{F}} \hat{\bm{\mu}} \big),
\end{aligned}
\end{equation}
with ${\bf{\mathcal{F}}} \in \mathbb{R}^{n_{0} N \times n_{o}} = \bm{1}_{N} \otimes {\bf{I}}_{n_{o}} $ where $\bm{1}_{N} \in \mathbb{R}^{N}$ is a vector of ones and  ${\bf{I}}_{n_{0}} \in \mathbb{R}^{n_{0} \times n_{0}}$ is a unit matrix. Furthermore,
$\boldsymbol{{\Pi}} \in \mathbb{R}^{n_{o} \times n_{o} N}$ and $\hat{\bm{\mu}} \in \mathbb{R}^{n_{o}}$ are defined as: 
\begin{subequations}
\begin{align}
& \boldsymbol{{\Pi}}({\bm{x}}_{\star}) = \begin{bmatrix}
Cov({\bm{Y}}({\bm{x}}_{\star}), {\bm{Y}}({\bm{c}}_{1})) & \cdots & Cov({\bm{Y}}({\bm{c}}_{\star}), {\bm{Y}}({\bm{c}}_{N})) \end{bmatrix}\, ,\\
& \hat{\bm{\mu}}= ({\bf \mathcal{F}}^{T} \boldsymbol{\Sigma}^{-1} {\bf \mathcal{F}})^{-1} {\bf \mathcal{F}}^{T} \boldsymbol{\Sigma}^{-1} {\bm{y}}^{tp}\, .
\end{align}
\end{subequations}

The prediction output of eq. (\ref{eq:mean}) is dependent on the values of the unknown trainable parameters $\bm{A}$ and ${\bm{\theta}}$. To simplify this procedure we assume in the following that the outputs are uncorrelated which means $\bm{A}$ is a-priori defined as a unit matrix.
The remaining parameters ${\bm{\theta}}$ can be found using a using a restricted maximum likelihood approach \citep{svenson2010multiobjective}
\begin{equation} \label{eq:optim_ML}
\begin{aligned}
\hat{{\bm{\theta}}} = \argmax_{ {\bm{\theta}}^{\star}} &\left[-\frac{1}{4} \log(|\bm{\Sigma}|) \log({\bf \mathcal{F}}^{T} \bm{\Sigma}^{-1} {\bf \mathcal{F}}) + \right. \\
&\left.-\frac{1}{2} ({\bm{y}}^{tp} - {\bf \mathcal{F}} \hat{\bm{\mu}})^{T} \bm{\Sigma}^{-1} ({\bm{y}}^{tp} - {\bf \mathcal{F}} \hat{\bm{\mu}}) \right]\, .
\end{aligned}
\end{equation}
After finding the best parameters, the GPR regression model is fully defined and predictions can be obtained by using equation (\ref{eq:mean}). 
\subsubsection*{Local approximate Gaussian process regression} 
Due to its setup as a nonparametric model, GPR suffers from computational intractability in the big data domain ($>1000$ data points) \cite{rasmussen2003gaussian}. Based on the works of \cite{gramacy2015local} and \cite{kleijnen2020prediction}, the authors in their recent work \citep{fuhg2021local} introduced local approximate GPR (laGPR) to the field of data-driven constitutive models. This technique keeps the major advantages of the general GPR method but makes it tractable for larger datasets. 
The basic concept is build around the premise that points closer to an input of interest $\bm{x}_{\star}$ have more influence to its output prediction than points far away from it.
This can be understood when observing the properties of the autocorrelation function of eq. (\ref{eq:autocorr}), if the elementwise difference between two points is sifgnificant the exponential will result in $R(\bm{x}_{\star}, \bm{x}', \bullet) = 0$. Other autocorrelation functions show similar effects, see e.g. \citep{fuhg2019adaptive}.
The idea in laGPR is to find a subset of cardinality $n \ll N$ of the whole dataset with $N$ training points with which a locally accurate surrogate model can be obtained. These input points are called inducing points.
Different techniques to find the inducing point set have been proposed and investigated in the literature 
Different variations of formulations for the inducing point set $\mathcal{X}_{n}$ have been explored and tested in the literature
, see \citep{gramacy2016lagp}.\\

Specifically, \cite{kleijnen2020prediction} simply use the $n$ nearest neighbors of the point $\bm{x}_{\star}$ in the whole dataset as measured by the euclidean distance
\begin{equation}
    d(\bm{x}_{\star}, \bm{x}) = \sqrt{(\bm{x}_{\star}- \bm{x})^{T} (\bm{x}_{\star}- \bm{x})}.
\end{equation}
Recently, it was shown by the authors that this formulation is able to accurately predict complex constitutive relationships based on data \citep{fuhg2021local}.
One negative side effect is that each output prediction requires a retraining of the unknown parameters of the local GPR model. However, as pointed out in \cite{kleijnen2020prediction}, even with around $100$ points in the local dataset the prediction can basically be done in real-time when choosing an efficient optimization algorithm.

\section{Numerical tests}\label{sec::4}
This section compares the presented physics-informed surrogate modeling approach for isotropic and anisotropic materials with the classical mapping approach. Furthermore, the efficiency of the proposed space-filling sampling approach is highlighted.
For all the following numerical tests consider the training domain to be $17.5 \%$ which means
\begin{equation}\label{eq:NumeDeforSpace}
        \overline{F}_{ij} \in [F^{L}_{ij}, F^{U}_{ij}] \text{ where } \begin{cases}
        0.825 \leq 1 \leq 1.175, & \text{when } i=j \\
        -0.175 \leq 0 \leq 0.175, & \text{when } i\neq j
        \end{cases}.
\end{equation}
For testing, we randomly sample $N_{t} =20,000$ points in this nine-dimensional training domain using LHS.
With reference to these points we define the mean stress output error as
\begin{equation}\label{eq::ErrorS}
    \mathcal{E}_{S} = \sqrt{\frac{1}{6 N_{t}}  \sum_{i=1}^{6} \sum_{j=1}^{N_{t}} \abs{\hat{\bm{s}}_{i}^{j} - \bm{s}_{i}^{j}}^{2}}
\end{equation}
where $\bm{s} = [S_{11}, S_{12}, S_{13}, S_{22}, S_{23}, S_{33}]^{T} $. \\

All results were obtained with MATLAB \citep{MATLAB:2021} \footnote{After acceptance of the paper the codes of this manuscript will be released under \url{https://github.com/FuhgJan/invariant_DD_CM}.  Python versions of the code can be made available under reasonable request.}. 
The utilized local approximate GPR model is a modified version of the code provided by \cite{lophaven2002dace}. 
In order to showcase the efficiency of the presented approaches we deliberately choose the isotropic and anistropic stress outputs to be on span different orders of magnitude. As described in Section \ref{sec::Sampling} evenly-spread points in the deformation gradient space might have the same exact invariants and would therefore not yield any new information for the training process. Even more problematic is that the covariance matrix of GPR and therefore of laGPR gets ill conditioned when two points in the training dataset are too close to each other \citep{fuhg2019adaptive}. Hence, when obtaining input points in the invariant space from the deformation gradient we remove duplicate points from the dataset. We define two points $\bm{x}^{1} \in \mathbb{R}^{d}$ and $\bm{x}^{2} \in \mathbb{R}^{d}$ as being a duplicate if 
\begin{equation}\label{eq::duplicate}
    \abs{\bm{x}^{1}_{i}-\bm{x}^{2}_{i}} < 0.01, \qquad \forall \, i=1, \ldots, d.
\end{equation}
This results in the fact that the size of the dataset used to train the classical mapping approach is always smaller or equal than the one for the physics-informed mappings of eq. (\ref{eq::IsoMapping}) and (\ref{eq:transIMapping}).

\subsection{Isotropic example}
Consider the compressible Mooney-Rivlin model of the form \citep{holzapfel2000nonlinear} 
\begin{equation}
    \Psi = c (J-1)^{2} - 2(c_{1}+c_{2})\ln{J} + c_{1} (I_{1}-3) + c_{2} (I_{2}-3)
\end{equation}
which yields the stress
\begin{equation}
\bm{S} = 2(c_{1} + c_{2} I_{1}) \bm{I} - 2 c_{2} \bm{C} + (2 c J (J-1)-2(c_{1}+c_{2})) \bm{C}^{-1}
\end{equation}
where we choose $c_{1}=c_{2}=1$, $c_{2}=0.2$ and $c_{3}=0.8$.
In the following we use this analytical model to test the performance of the presented physics-informed surrogate modeling approach for the isotropic case with and without space-filling sampling and compare the results to the classical mapping approach as described in eq. (\ref{eq:naiveMap}).

\begin{figure}[h]
    \centering
    \includegraphics[scale=0.5]{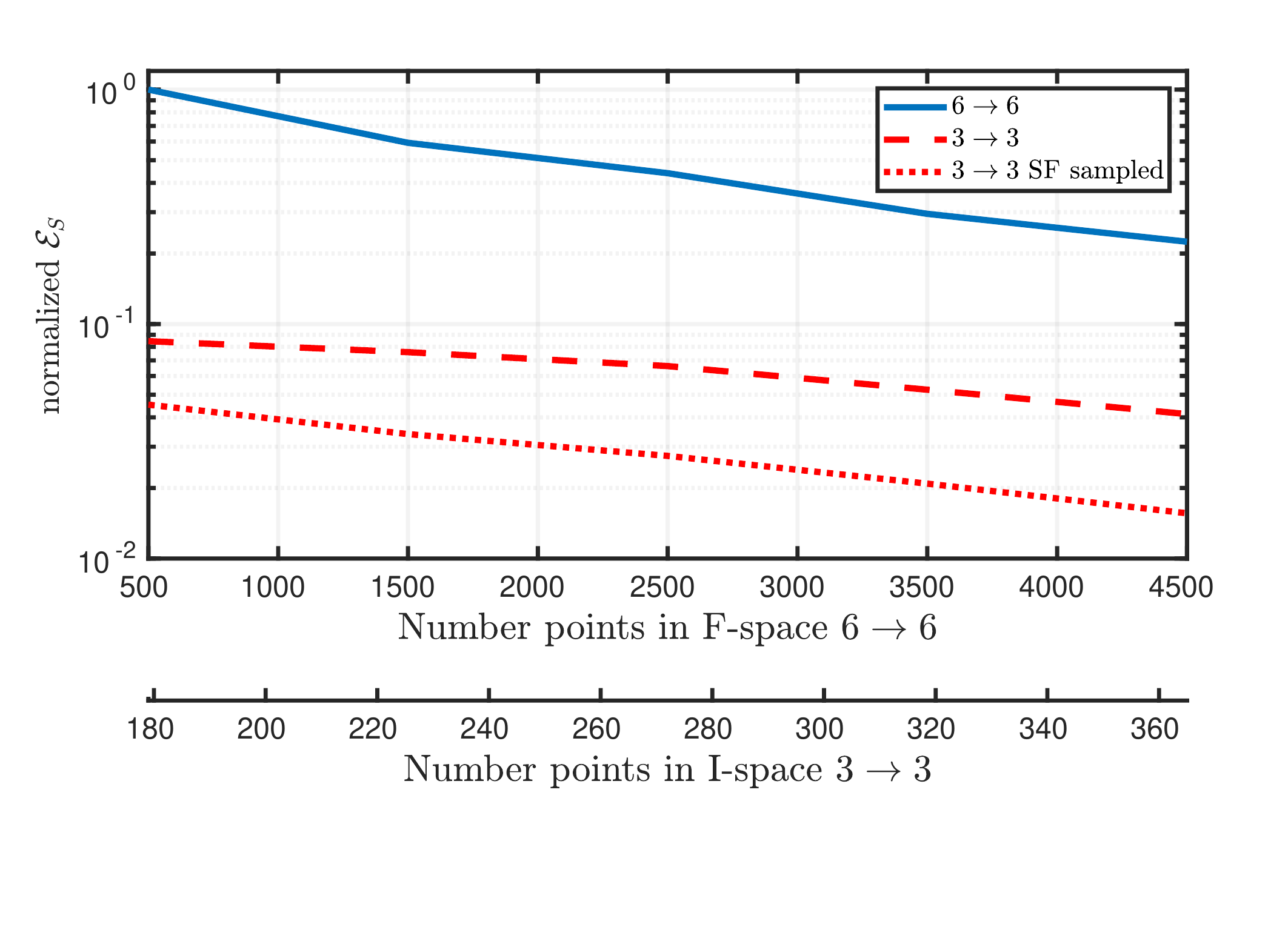}
    \caption{Normalized error values for isotropic hyperelastic law using $20,000$ test points in input range of $17.5\%$ sampled with LHS. Three mapping approaches: the classical mapping approach $(6 \rightarrow 6)$, mapping in the invariant space $(3 \rightarrow 3)$ and mapping in the invariant space where the input points have been sampled with the proposed space-filling technique $(3 \rightarrow 3 \text{ SF sampled})$. Points have been sampled in deformation gradient space for $6 \rightarrow 6$ and then projected onto invariant space which reduced the number of points. Space-filling sampling has been done with the number of points in invariant space.}
    \label{fig:isoErrors}
\end{figure}

For all the following results the classical mapping was trained with 
$60$ inducing points. Since for all the investigated cases the number of points for the physical-mapping never exceeded $360$ standard GPR was used for all of the following cases.
Figure \ref{fig:isoErrors} shows the error for the $20,000$ randomly sampled points with LHS for the classical mapping approach ($6 \rightarrow 6$), the physics-informed mapping approach ($3 \rightarrow 3$) and the physics-informed mapping approach where the inputs are generated with the space-filling sampling technique of Algorithm \ref{algo::EvenIso} ($3 \rightarrow 3$ SF sampled). 
For ($6 \rightarrow 6$), between $500$ and $4500$ points are sampled in the nine-dimensional deformation gradient space of eq. (\ref{eq:NumeDeforSpace}) with TPLHD. Using eq. (\ref{eq::duplicate}) the resulting invariant space datasets are then checked for duplicates, which results in datasets of sizes ranging between $180$ and $360$ samples. These samples are used to train the ($3 \rightarrow 3$) approach. Using the same number of points as for ($3 \rightarrow 3$) the Algorithm \ref{algo::EvenIso} is used to generate space-filling samples which were used to build the datasets for ($3 \rightarrow 3$ SF sampled).
From Figure \ref{fig:isoErrors} it can be seen that even though the surrogate models in invariant space were trained on vastly lower number of points they perform better by around a factor $10$ compared to the classical mapping approach. Additionally, the datasets that were created with the space-filling algorithm show large improvements in comparison to the plain ($3 \rightarrow 3$) approach which uses samples that were generated in the deformation gradient space.

\begin{figure}[ht]
\begin{subfigure}[b]{0.5\linewidth}
\centering
\includegraphics[scale=0.35]{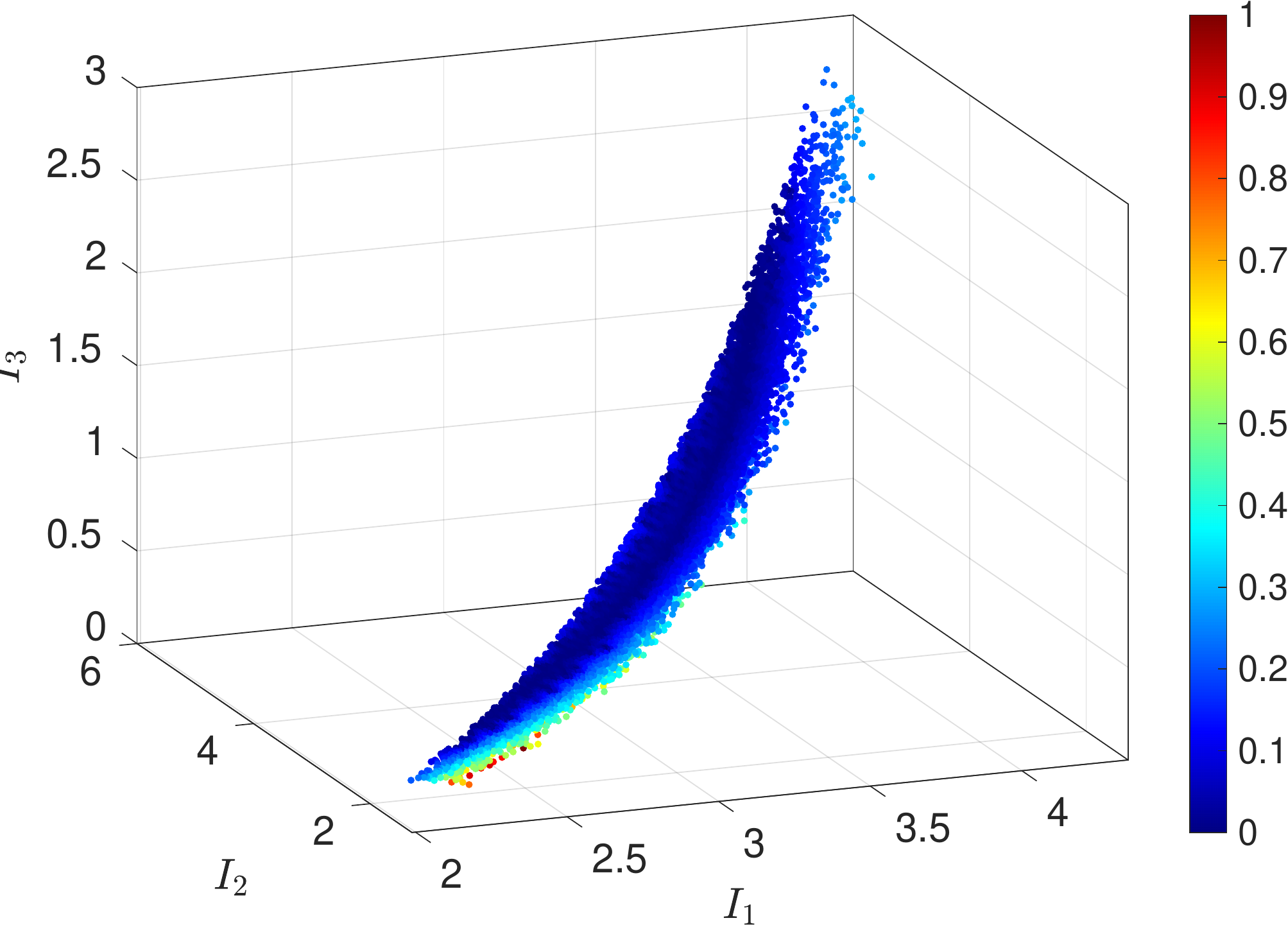} 
\caption{$275$ samples obtained from deformation gradient space}\label{fig:isoErrorsIn3Da}
\end{subfigure}%
\begin{subfigure}[b]{.5\linewidth}
\centering
\includegraphics[scale=0.35]{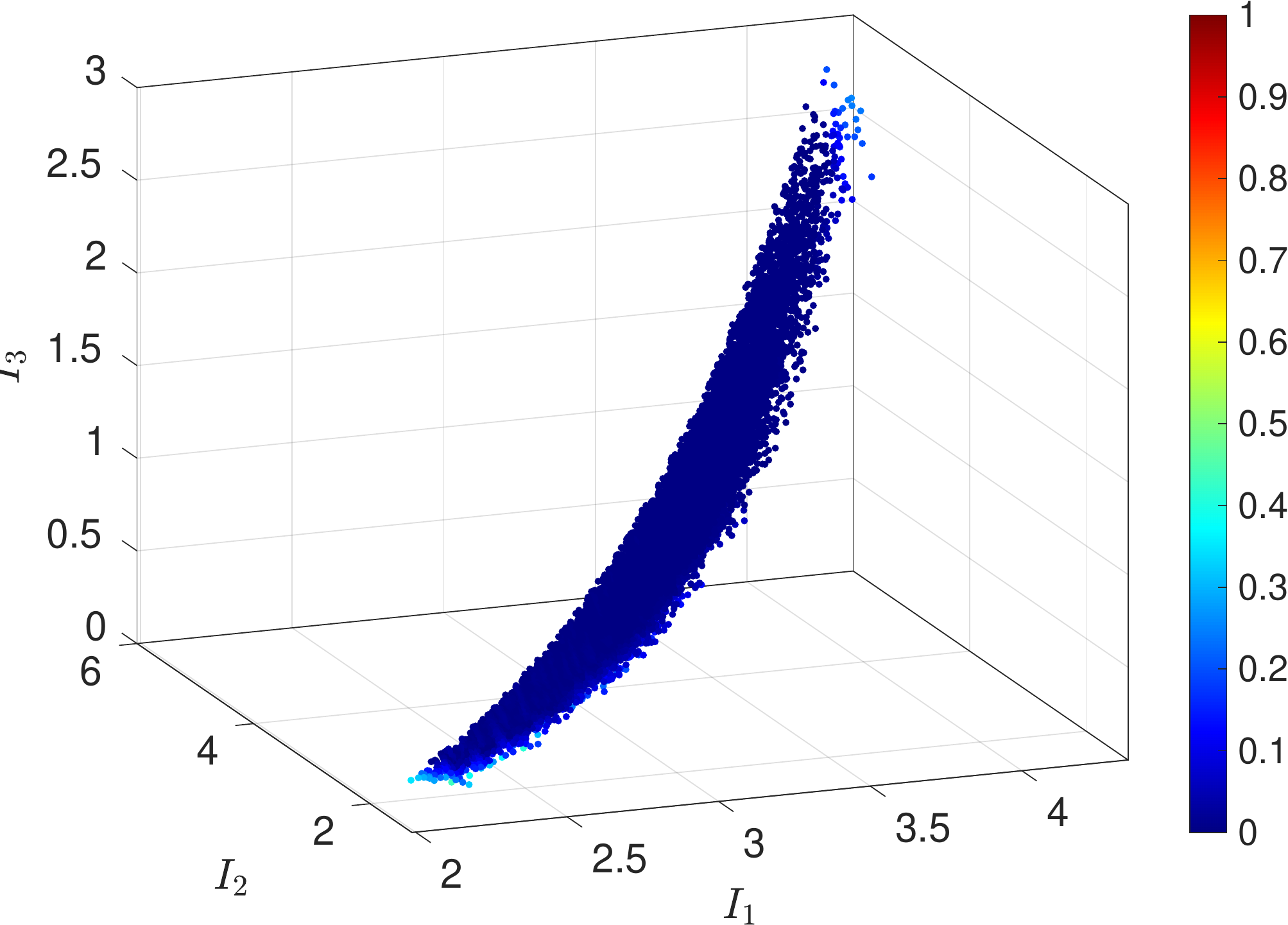} 
\caption{$275$ samples generated with space-filling approach}\label{fig:isoErrorsIn3Db}
\end{subfigure}
\caption{Normalized error over $20,000$ test points in $17.5\%$ input range surrogate models trained with $275$ sample points (a) $2,500$ points (sampled with TPLHD) in deformation gradient space and then projected onto invariant space which results in $275$ samples when reducing duplicates ($3 \rightarrow 3$), (b) $278$ sampled with the space-filling sampling approach of Algorithm \ref{algo::EvenIso} ($3 \rightarrow 3$ SF sampled).}\label{fig:isoErrorsIn3D}
\end{figure}

The differences between ($3 \rightarrow 3$) and ($3 \rightarrow 3$ SF sampled) are further highlighted in
Figure \ref{fig:isoErrorsIn3D} which plots the normalized error values in invariant space for the $20,000$ test points and $2500$ deformation gradient training points ($275$ training points in invariant space) and which emphasizes the location of the largest errors. As expected, the areas of the largest errors of the ($3 \rightarrow 3$) approach (Figure \ref{fig:isoErrorsIn3Da}) are around the boundaries of the test data domain since the invariant is not sampled evenly. The errors of the space-filling dataset as shown in Figure \ref{fig:isoErrorsIn3D} are uniformly lower and especially the boundary regions are better represented.

\begin{figure}[h]
\begin{subfigure}[b]{0.5\linewidth}
\centering
\includegraphics[scale=0.35]{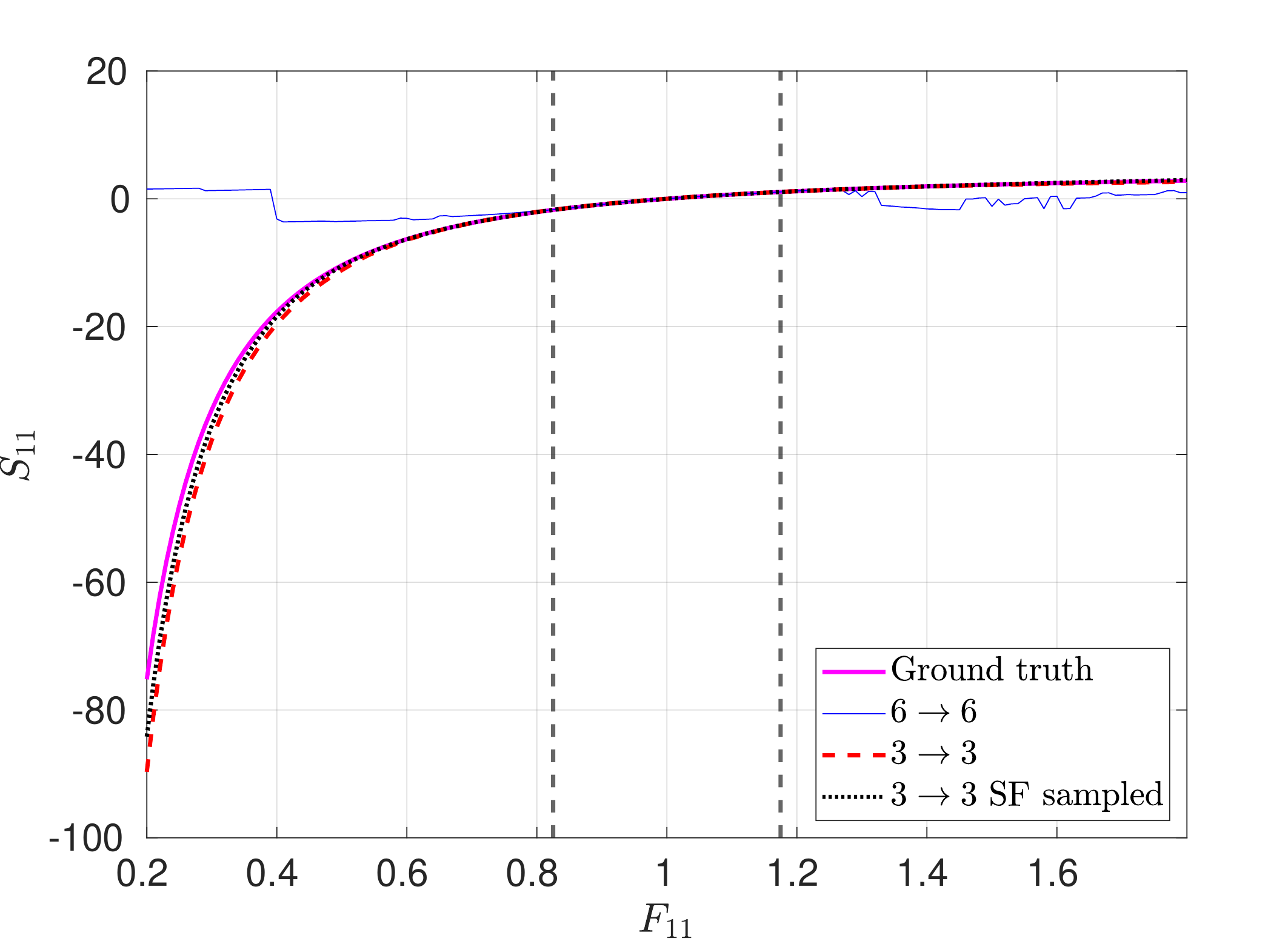} 
\caption{Real and predicted $S_{11}$ over $F_{11}$}\label{fig::IsoExamplestressOutputsa}
\end{subfigure}%
\begin{subfigure}[b]{.5\linewidth}
\centering
\includegraphics[scale=0.35]{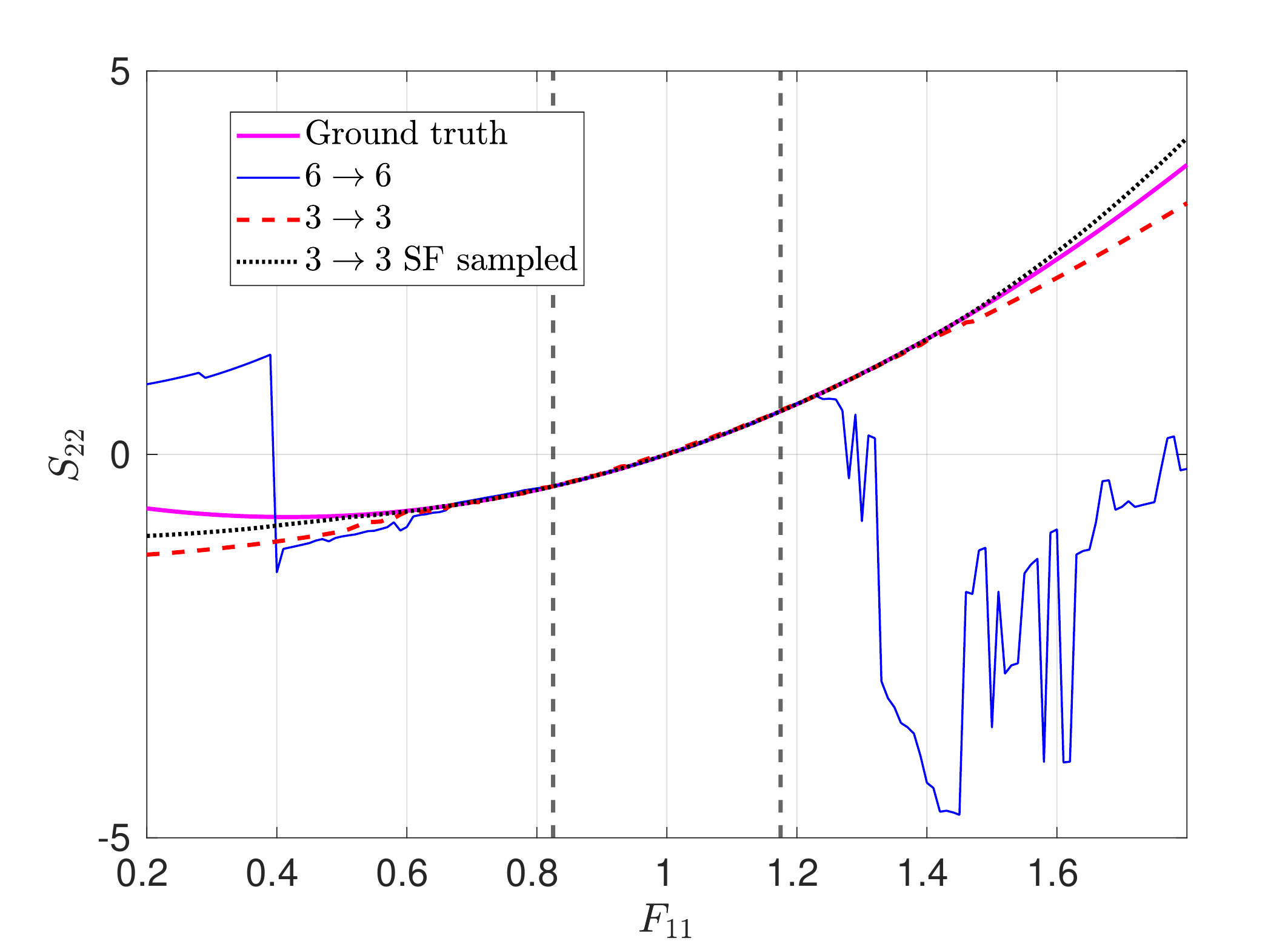} 
\caption{Real and predicted $S_{12}$ over $F_{11}$}\label{fig::IsoExamplestressOutputsb}
\end{subfigure}
\begin{subfigure}[b]{0.5\linewidth}
\centering
\includegraphics[scale=0.35]{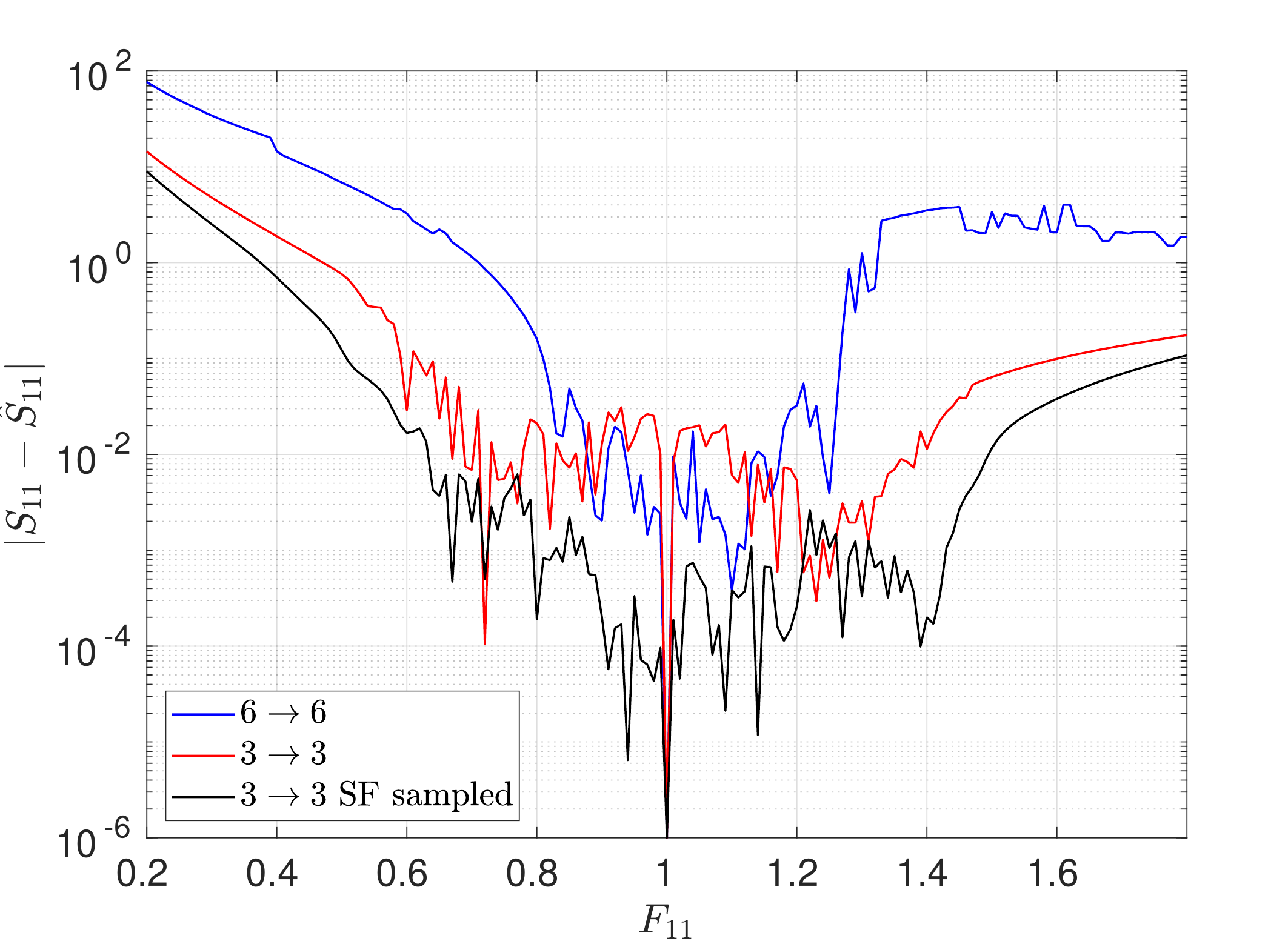} 
\caption{$|S_{11}-\hat{S}_{11}|$ over $F_{11}$}\label{fig::IsoExamplestressOutputsc}
\end{subfigure}%
\begin{subfigure}[b]{.5\linewidth}
\centering
\includegraphics[scale=0.35]{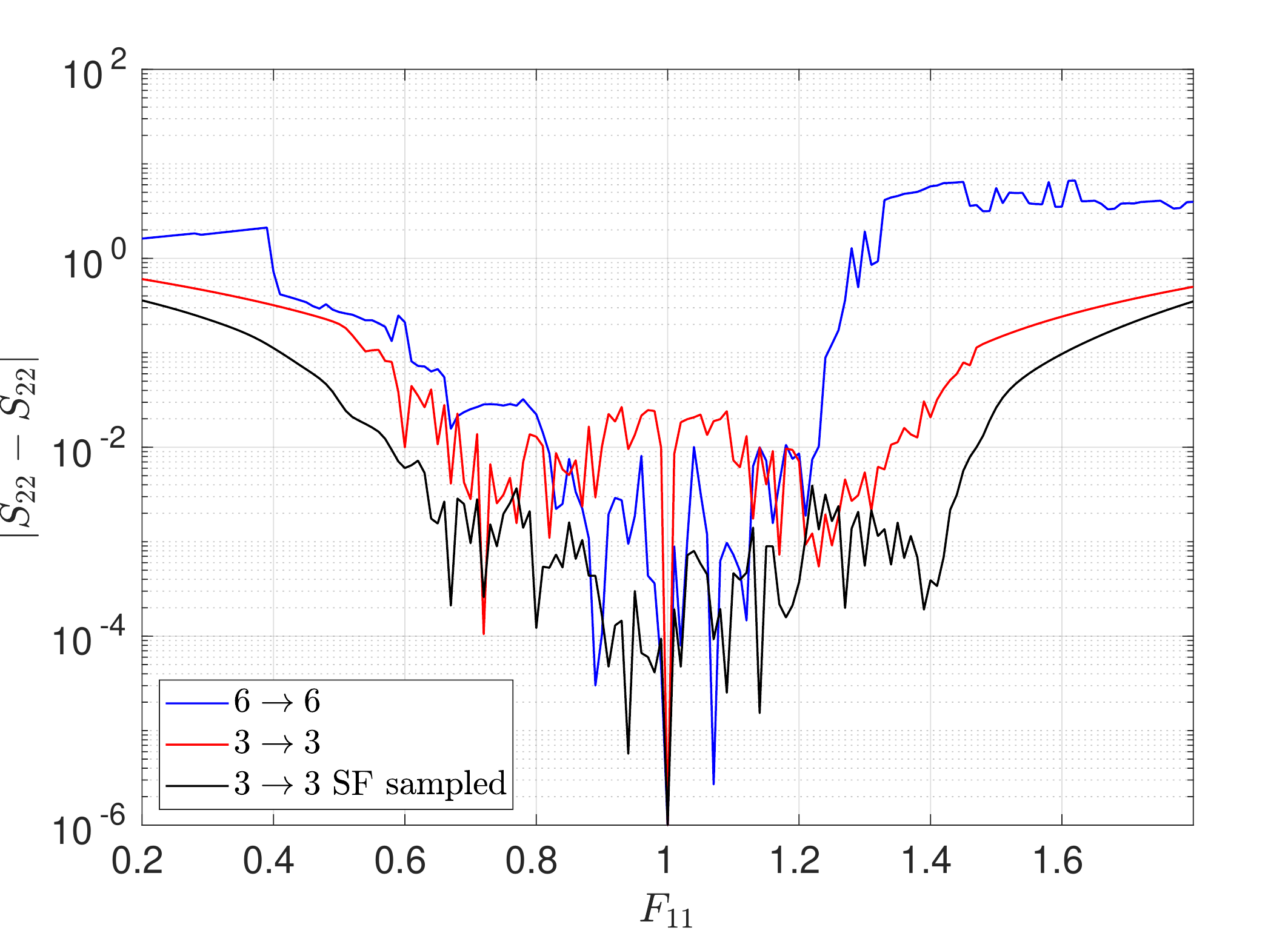} 
\caption{$|S_{22}-\hat{S}_{22}|$ over $F_{11}$}\label{fig::IsoExamplestressOutputsd}
\end{subfigure}
\caption{Illustrative stress outputs for the isotropic case. Thicker horizontal dashed lines symbolize positions of $17.5\%$ training domain.}\label{fig::IsoExamplestressOutputs}
\end{figure}

Consider the following applied deformation gradient range
\begin{equation}
    \bm{F}_{app} =\bm{I} + F_{11,app} \bm{e}_{1} \otimes \bm{E}_{1}, \qquad \text{with }   F_{11,app} \in [-0.8, 0.8]
\end{equation}
not explicitly part of the training dataset, and which in fact applies loads far outside the input training domain of eq. (\ref{eq:NumeDeforSpace}).
The ground truth responses to this load in directions $S_{11}$ and $S_{22}$ as well as the predicted responses using $2500$ TPLHD points for ($6 \rightarrow 6$) and $278$ points for ($3 \rightarrow 3$) and ($3 \rightarrow 3$ SF sampled) after removing the duplicates are shown in Figures  \ref{fig::IsoExamplestressOutputsa} and \ref{fig::IsoExamplestressOutputsb}. The corresponding absolute errors between the predicted and actual responses for these two cases are shown in Figures \ref{fig::IsoExamplestressOutputsc} and \ref{fig::IsoExamplestressOutputsd} respectively. It can be seen that all of the investigated approaches are able to accurately capture the stress response within the training region
(marked by the dotted lines in Figures \ref{fig::IsoExamplestressOutputsa} and \ref{fig::IsoExamplestressOutputsb}). Notably, outside of the training domain the physics-informed approaches perform with significantly better accuracy compared to the classical mapping counterpart. Additionally, the space-filling approach shows better results than the normal ($3 \rightarrow 3$) mapping. This is the first example where physics-informed data-driven constitutive modeling in the context of hyperelasticity is shown to generalize proficiently. The initial training region corresponds to $17.5\%$ but the results of Figures \ref{fig::IsoExamplestressOutputsa} and \ref{fig::IsoExamplestressOutputsb} are shown in a testing region of $80\%$. Interestingly, it can be seen that the classical approach, which is solely data-driven provides non-physical results beyond the $17.5\%$ training region.

\subsection{Transversely isotropic example}
\begin{figure}[h]
    \centering
    \includegraphics[scale=0.5]{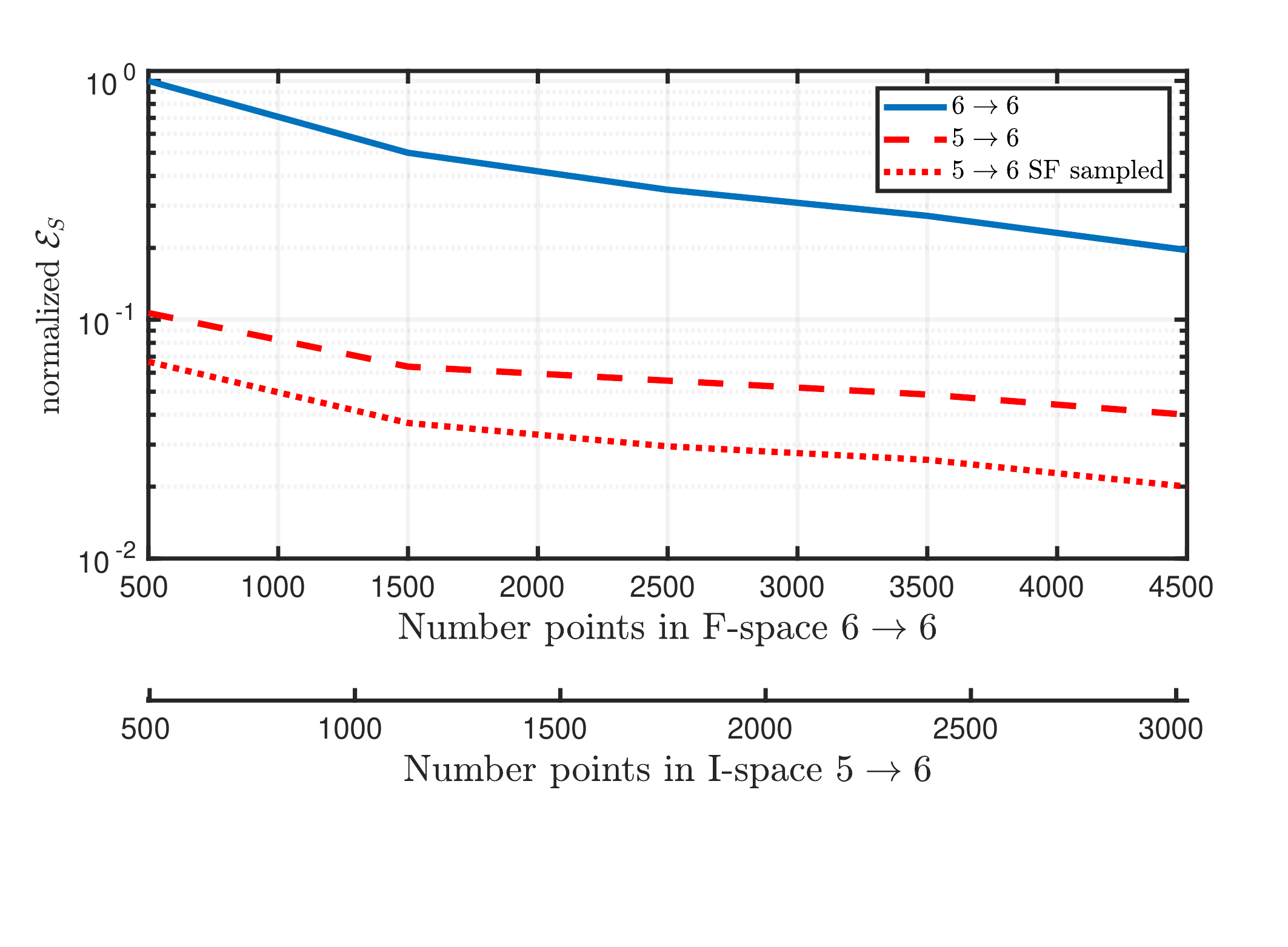}
    \caption{Normalized error values for anisotropic hyperelastic law using $20,000$ test points in input range of $17.5\%$ sampled with LHS. Three mapping approaches: the classical mapping approach $(6 \rightarrow 6)$, mapping in the invariant space $(5 \rightarrow 6)$ and mapping in the invariant space where the input points have been sampled with the proposed space-filling technique $(5 \rightarrow 6 \text{ SF sampled})$. Points have been sampled in deformation gradient space for $6 \rightarrow 6$ and then projected onto invariant space which reduced the number of points. Space-filling sampling has been done with the number of points in invariant space.}
    \label{fig:transIsoErrors}
\end{figure}

\begin{figure}[ht]
\begin{subfigure}[b]{0.5\linewidth}
\centering
\includegraphics[scale=0.35]{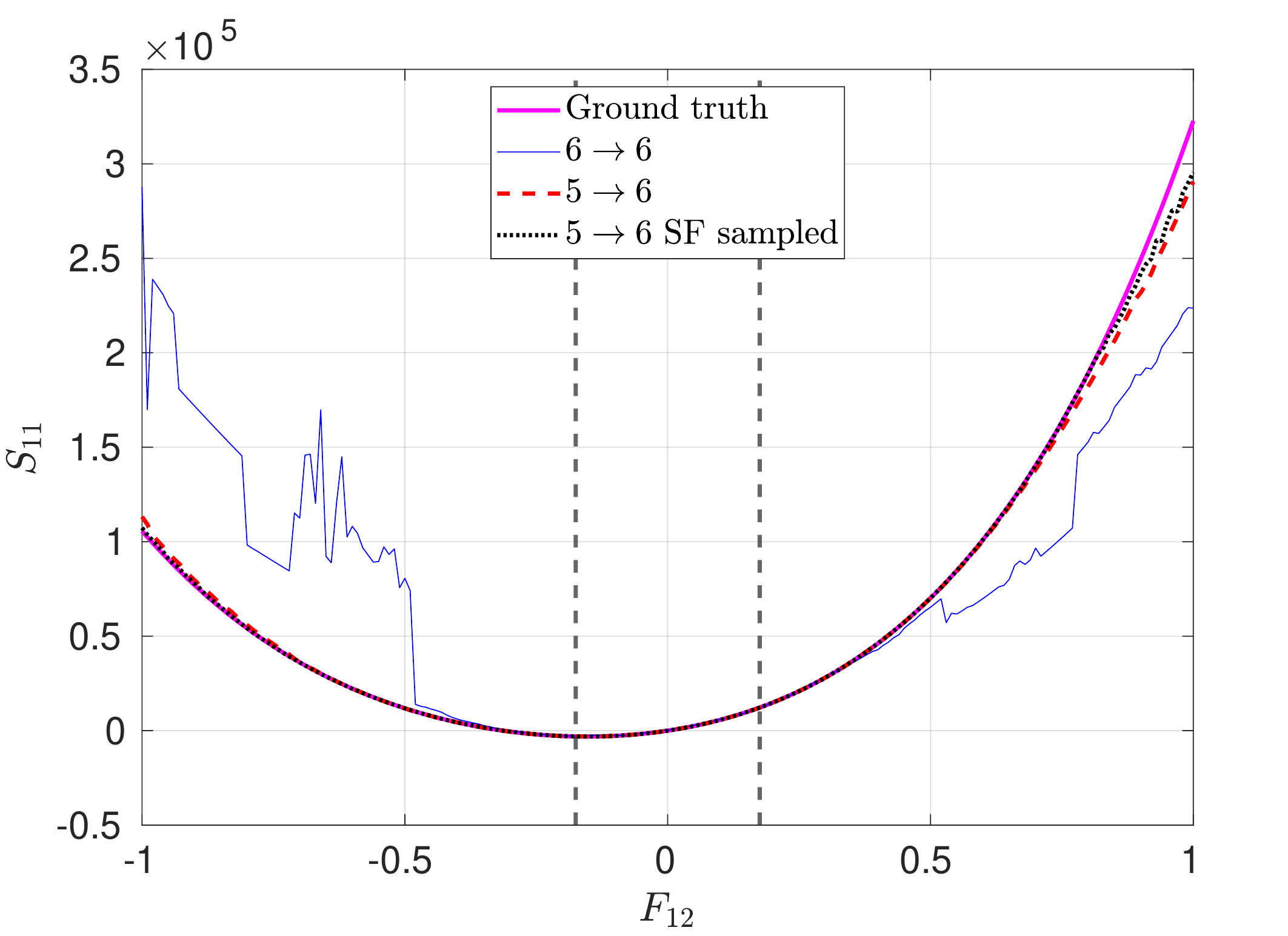} 
\caption{Real and predicted $S_{11}$ over $F_{12}$}\label{fig::TransIsoStressPlota}
\end{subfigure}%
\begin{subfigure}[b]{.5\linewidth}
\centering
\includegraphics[scale=0.35]{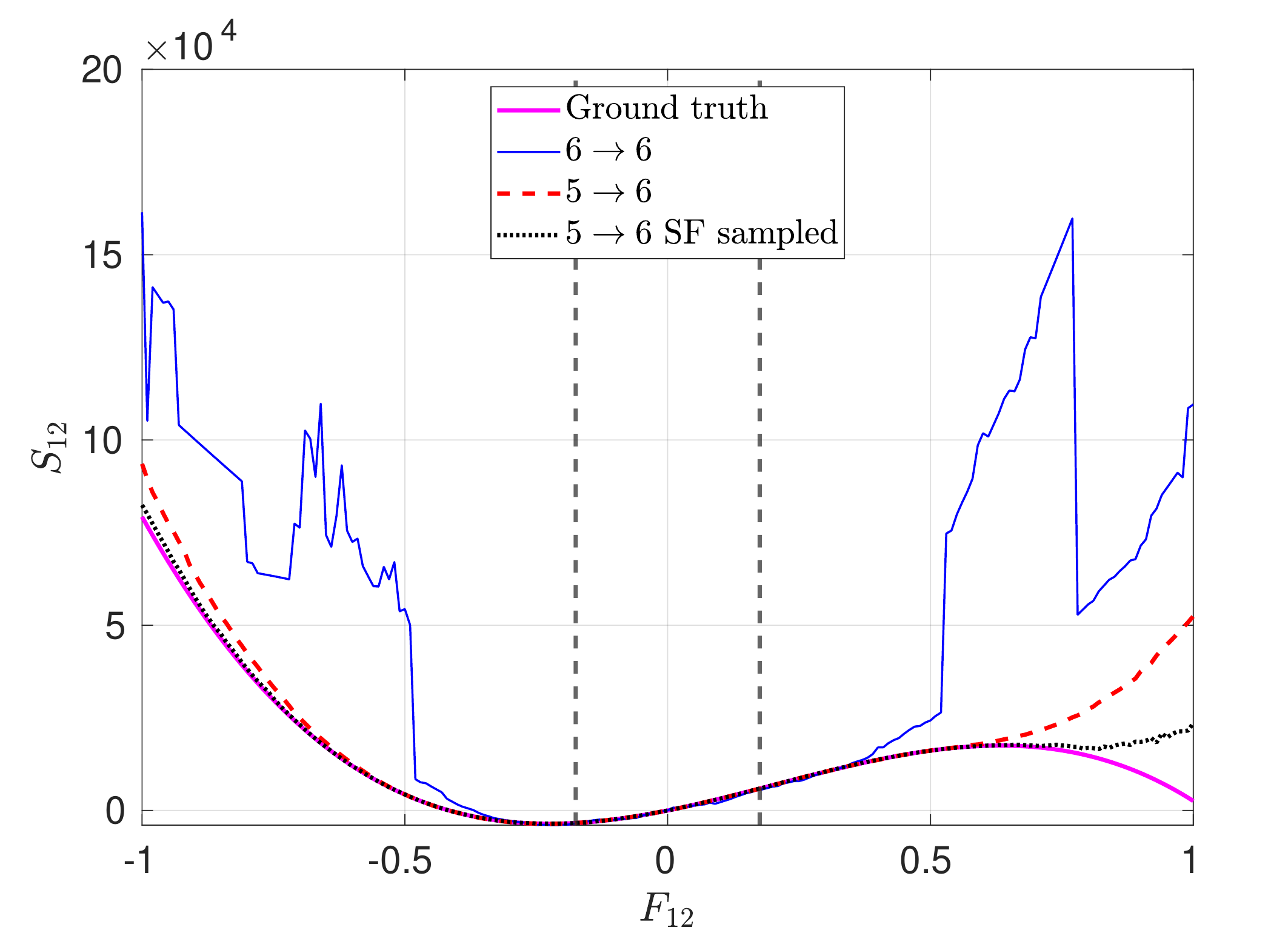} 
\caption{Real and predicted $S_{12}$ over $F_{12}$}\label{fig::TransIsoStressPlotb}
\end{subfigure}
\begin{subfigure}[b]{0.5\linewidth}
\centering
\includegraphics[scale=0.35]{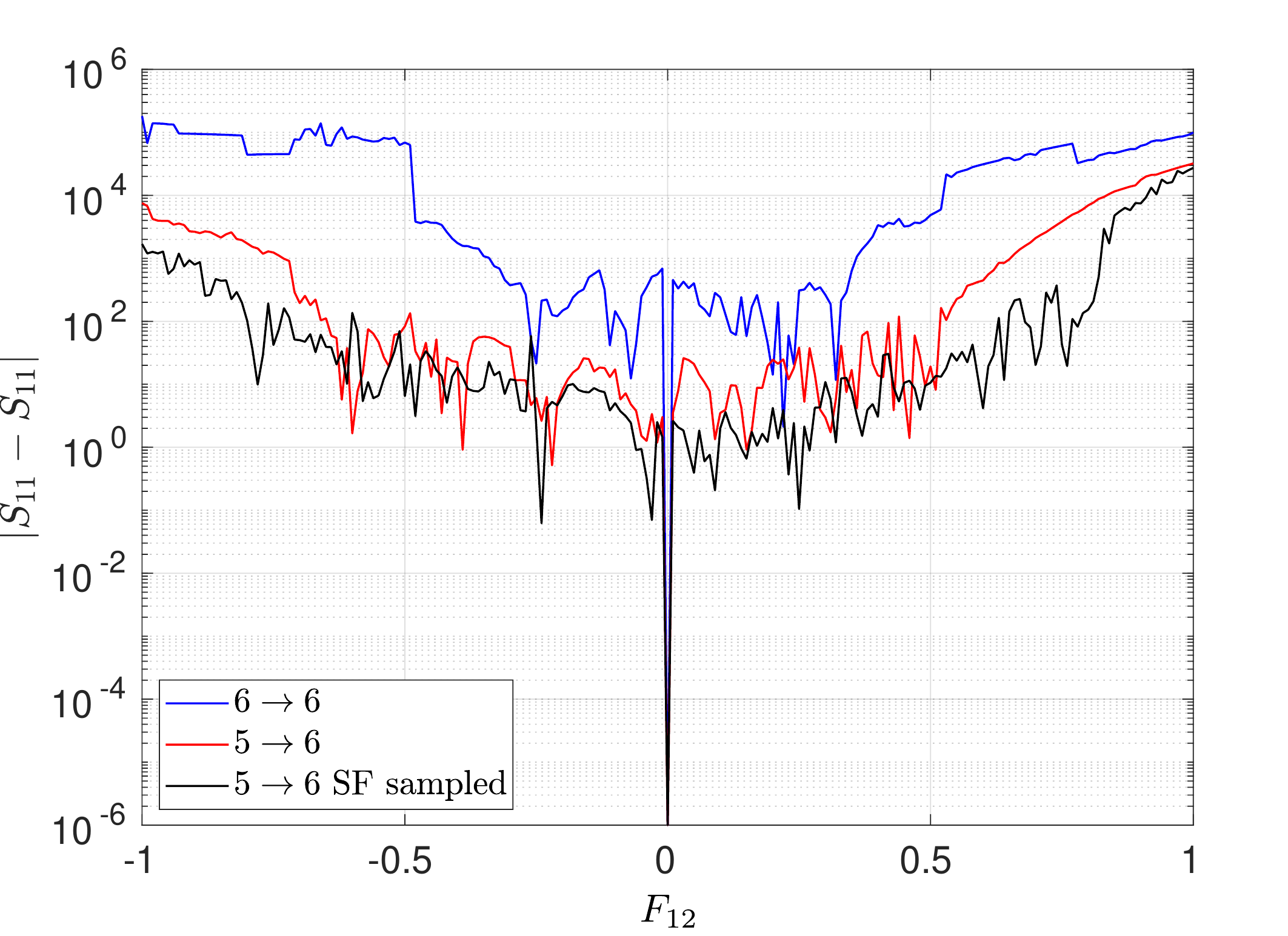} 
\caption{Real and predicted $S_{11}$ over $F_{12}$}\label{fig::TransIsoStressPlotc}
\end{subfigure}%
\begin{subfigure}[b]{.5\linewidth}
\centering
\includegraphics[scale=0.35]{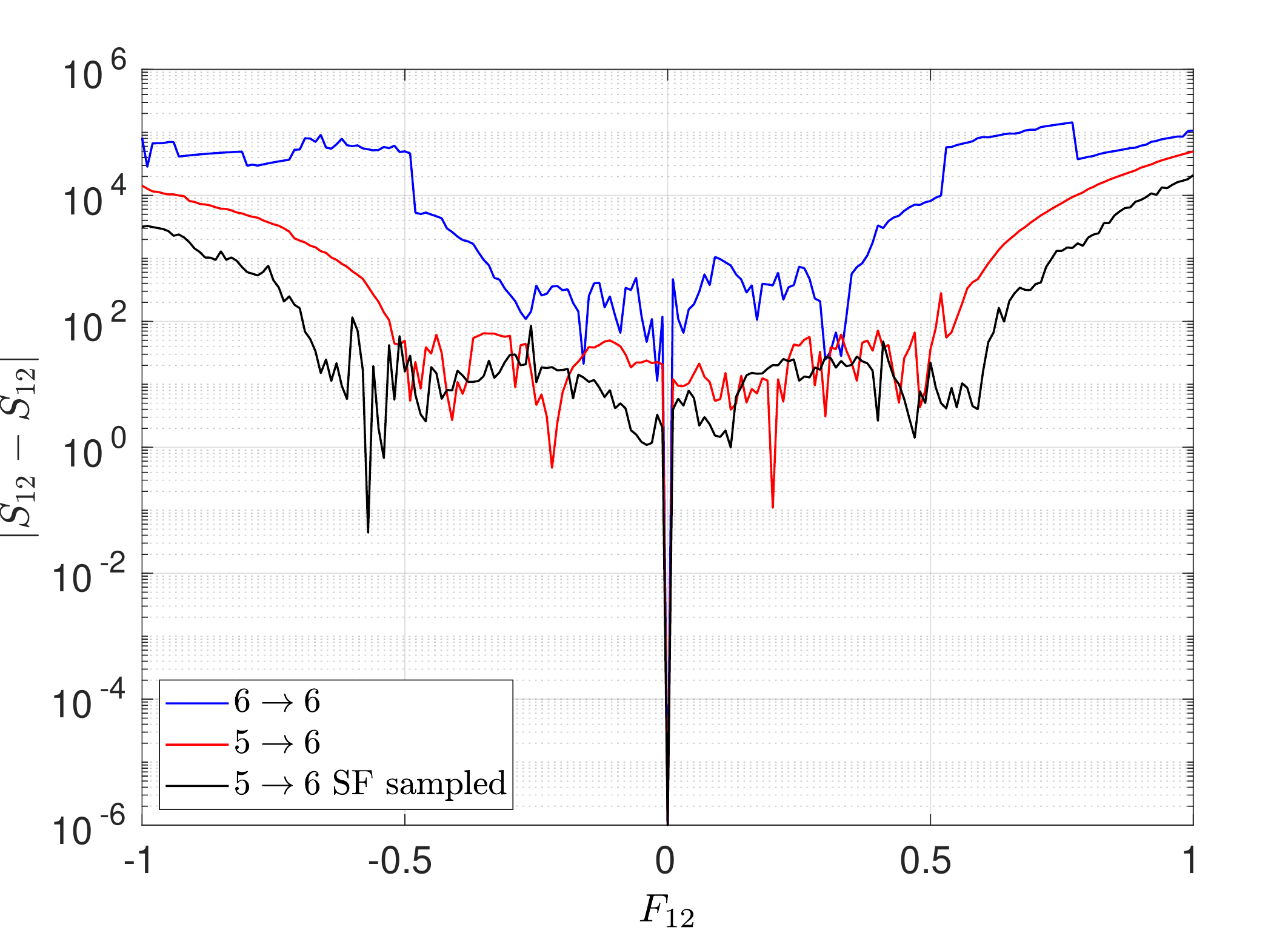} 
\caption{Real and predicted $S_{12}$ over $F_{12}$}\label{fig::TransIsoStressPlotd}
\end{subfigure}
\caption{Illustrative stress outputs for the anisotropic case. Thicker horizontal dashed lines symbolize positions of $17.5\%$ training domain.}\label{fig::TransIsoStressPlot}
\end{figure}

Consider the following analytical transversely isotropic hyperelastic law suggested by \cite{bonet1998simple} which is given as
\begin{equation}
    \Psi = [\alpha + \beta \log J + \gamma (I_{4}-1) ] (I_{4} - 1) - \frac{1}{2} \alpha (I_{5}-1) ,
\end{equation}
yielding the corresponding second Piola-Kirchhoff stress tensor
\begin{equation}
    \bm{S} = 2 \beta (I_{4}-1) \bm{C}^{-1} + 2 [\alpha + 2 \beta \log{J} + 2 \gamma (I_{4}-1)] \bm{a}_{0} \otimes \bm{a}_{0} - \alpha \left( \bm{C} \bm{a}_{0} \otimes \bm{a}_{0} + \bm{a}_{0} \otimes \bm{C} \bm{a}_{0}\right)
\end{equation}
where $\bm{a}_{0}= \frac{1}{\sqrt{6}}[1,2,1]^{T}$ is a unit vector representing the direction of reinforcement, $I_{4} = \bm{C}: \bm{a}_{0} \otimes \bm{a}_{0}$ the fourth principal invariant, and normalized shear, bulk and fiber reinforcement moduli of $\alpha=1.585e5$, $\beta=5e4$ and $\gamma=1.8e5$. 
This analytical model is used to test the performance of the presented physics-informed surrogate modeling approach for the anisotropic materials with and without space-filling sampling (here termed ($5 \rightarrow 6$) and ($5 \rightarrow 6$ SF sampled) respectively). These results are compared to the classical mapping approach ($6 \rightarrow 6$) as described in eq. (\ref{eq:naiveMap}). For all the following results all mappings were trained with laGPR and $60$ inducing points. 
Figure \ref{fig:transIsoErrors} shows the errors between the three cases for $20,000$ test points in the training domain of eq. (\ref{eq:NumeDeforSpace}) generated with LHS.
Similary to the isotropic case the mapping ($6 \rightarrow 6$) is trained on datasets consisting of $500$ to $4500$ points which are sampled in the nine-dimensional deformation gradient space of eq. (\ref{eq:NumeDeforSpace}) with TPLHD. These datasets are then checked for duplicates using eq. (\ref{eq::duplicate}) which reduces the size of the datasets for training from the invariant space to a range between $500$ and $3500$ samples. These define datasets which are in turn used to train the ($3 \rightarrow 3$) approach. An equivalent number of samples is then respectively generated using the Algorithm \ref{algo::EvenIso} to obtain datasets consisting of space-filling samples in primary and pseudo-invariant space which are used to train ($3 \rightarrow 3$ SF sampled). It can be seen that training the surrogate in a physics-informed way decreases the error dramatically while at the same time less training samples are needed. Here again, the datasets that were generated in a space-filling manner outperform the datasets that were sampled in deformation gradient space.\\
Figure \ref{fig::TransIsoStressPlot} shows raw stress and error output values when the following deformation gradients are applied
\begin{equation}
    \bm{F}_{app} = \bm{I} + F_{12,app} \bm{e}_{1} \otimes \bm{E}_{2}, \qquad \text{with }  F_{12,app} \in [-1, 1].
\end{equation}
This load path is
not explicitly part of the training dataset, and even applies loads far beyond the input training domain of eq. (\ref{eq:NumeDeforSpace}).
Figures \ref{fig::TransIsoStressPlota} and \ref{fig::TransIsoStressPlotb} show the ground truth stress responses as well the predicted stresses in directions $S_{11}$ and $S_{12}$. The respective absolute errors are shown in 
Figures \ref{fig::TransIsoStressPlotc} and \ref{fig::TransIsoStressPlotb}. 
The shown results are based on a  ($6 \rightarrow 6$) model trained with $2000$ training points and physics-informed models based on a corresponding $1480$ points. 
Similarly to the isotropic case, it can be seen that all the surrogate models are able to capture the response inside the training domain (dotted vertical lines) in a proficient way. However outside the training domain the classical mapping approach become crucially unreliable while the physics-informed metamodels are able to follow the true response surprisingly far away from the main training domain once again confirming their ability to efficiently generalize. This indicates that training models that guarantee material frame indifference, material symmetry and thermodynamic consistency allows them to intrinsically learn the involved physics of the material law. Here again, the space-filling sampling approach shows better performances than the model that was trained from the dataset that was sampled in the deformation gradient space.

\section{Discussion and outlook}\label{sec::5}
This paper presents a technique to obtain physics-informed data-driven surrogates for hyperelastic material models based on training data from isotropic and anisotropic materials. The idea is based on writing the second Piola-Kirchhoff stress as a linear combination of an irreducible basis of stress generators. This allows us to build metamodels that map from the corresponding invariants of the input to the scalars of the linear combination. The trained models are then able to inherently capture five physical concepts: the preservation of the stress-free undeformed configuration, local balance of angular momentum, material frame indifference, material symmetry conditions and thermodynamic consistency.
The surrogate modeling technique of choice is an approach called local approximate Gaussian process regression which in contrast to neural networks is a non-parametric model with convergence guarantees.

It is shown that the presented technique vastly outperforms the classical mapping approach which maps the symmetric right Cauchy-Green tensor components to the symmetric components of the second Piola-Kirchhoff stress tensor, without consideration of material frame indifference or material symmetry of the trained model. Surprisingly, the surrogates trained with the presented physics-informed concept were able to accurately capture stress paths which reached far outside the training domain showcasing their ability to generalize efficiently.
Furthermore a space-filling sampling technique is proposed that is able to generate evenly spread samples in the invariant space, for isotropic and anisotropic materials, based on some bounded deformation gradient domain. The sampling technique was explained in detail and its effectiveness in comparison to randomly obtained samples from the deformation gradient space was highlighted for isotropic and anisotropic numerical examples. 
In future works we aim to further study the surprising capabilities of the model to generalize even outside the training domain. This fact could allow us to generate accurate surrogate models with very sparse datasets spanning a large input domain.
\clearpage

\section*{Appendix}

\subsection*{List of tensor gradient expressions}
\begin{equation}\label{eq::ListOfGradientExpressions}
    \begin{aligned}
        \frac{\partial I_{1}}{\partial \bm{C}} &= \bm{I} \\
    \frac{\partial I_{2}}{\partial \bm{C}} &= I_{1}\bm{I} - \bm{C} \\
    \frac{\partial I_{3}}{\partial \bm{C}} &= I_{3} \bm{C}^{-1} \\
         \frac{\partial I_{4}}{\partial \bm{C}} &= \bm{A} \\
     \frac{\partial I_{5}}{\partial \bm{C}} &= \bm{a}_{0} \otimes \bm{C} \bm{a}_{0} +  \bm{a}_{0} \bm{C} \otimes \bm{a}_{0} \\
    \frac{\partial \bm{I}}{\partial \bm{C}} &= \bm{0} \otimes \bm{0} \\
    \frac{\partial \bm{C}}{\partial \bm{C}} &= \bm{I} \otimes \bm{I} \\
         \frac{\partial \bm{A}}{\partial \bm{C}} &= \bm{0} \otimes \bm{0} \\
     \left(\frac{\partial \bm{C}^{-1}}{\partial \bm{C}} \right)_{ijkl} &= - \frac{1}{2} \left( C_{ik}^{-1} C_{lj}^{-1} + A_{il}^{-1} A_{kj}^{-1} \right) \\
    \left( \frac{\partial  \bm{C}^{2}}{ \partial \bm{C}} \right)_{ijkl} &=  \delta_{ik} C_{lj} + C_{ik} \delta_{jl}\\
    \left( \frac{\partial  (\bm{A}\bm{C} + \bm{C}\bm{A})}{ \partial \bm{C}} \right)_{ijkl}  &= A_{ik} \delta_{jl} + \delta_{ik} A_{lj} \\
      \left( \frac{\partial  (\bm{A}\bm{C}^{2} + \bm{C}^{2}\bm{A})}{ \partial \bm{C}} \right)_{ijkl} &= A_{ik} C_{lj} + A_{im} C_{mk} \delta_{jl} + \delta_{ik} C_{lr} A_{rj} + C_{ik} A_{lj}
    \end{aligned}
\end{equation}

\bibliography{bib.bib}
\end{document}